\documentclass[aps,pre,reprint,showpacs,twocolumn,amsmath,amssymb,10pt,floatfix]{revtex4-1}

\usepackage{graphicx}
\usepackage{color}
\usepackage{hyperref}

\newcommand{\ket}[1]{|\,{#1}\,\rangle}

\newcommand{\rref}[1]{Ref.~\onlinecite{#1}}
\newcommand{\rrefs}[1]{Refs.~\onlinecite{#1}}
\newcommand{\fref}[1]{Fig.~\ref{#1}}
\newcommand{\frefp}[2]{Fig.~\ref{#1}~#2}
\newcommand{\eref}[1]{Eq.~(\ref{#1})}
\newcommand{\esref}[2]{Eqs.~(\ref{#1}) and (\ref{#2})}

\newcommand{\pref}[1]{(\ref{#1})}

\begin{document}

\title{Pseudomodes and the corresponding transformation of the temperature-dependent bath correlation function}

\author{David~W.~Sch{\"o}nleber}
\author{Alexander~Croy}
\author{Alexander~Eisfeld}
\email{eisfeld@pks.mpg.de}
\affiliation{Max Planck Institute for the Physics of Complex Systems, N\"othnitzer Strasse 38, 01187 Dresden, Germany}

\begin{abstract}
In open system approaches with non-Markovian environments, the process of inserting an individual mode (denoted as ``pseudomode'') into the bath or extracting it from the bath is widely employed. 
This procedure, however, is typically performed on basis of the spectral density (SD) and does not incorporate temperature.
Here, we show how the -- temperature-dependent -- bath correlation function (BCF) transforms in such a process. We present analytic formulae for the transformed BCF and numerically study the differences between factorizing initial state and global thermal (correlated) initial state of mode and bath, respectively. We find that in the regime of strong coupling of the mode to both system and bath, the differences in the BCFs give rise to pronounced differences in the dynamics of the system.
\end{abstract}


\maketitle


\section{Introduction}\label{sec:intro}
In many open quantum system approaches, the microscopic model underlying the environment consists of an infinite number of harmonic oscillators linearly coupled to system degrees of freedom \cite{weiss2008,may2011,mukamel1995}.
The flexibility of this particular model is owed to the fact that both spectrum and frequency-dependent coupling of the environmental modes can be adjusted to reproduce features observed in experiments, 
e.g.\ to describe the effect of polar solvents on dyes \cite{may2011,mukamel1995,fleming1996} or to treat vibrational modes of molecules \cite{may2011,roden2012}.  
The spectrum and the frequency-dependent coupling can be encoded in a quantity called spectral density (SD), which is a real function of frequency.

The influence of the environment on the system is eventually determined by the bath correlation function (BCF) \cite{may2011,weiss2008,diosi1998,roden2011,suess2014}, which quantifies temporal correlations of environmental degrees of freedom. Via the initial state of the environment, environment temperature is incorporated in the BCF.

The partitioning of the total system into explicitly treated degrees of freedom and a rest which is treated by means of a BCF is not fixed in the first place, and different choices might be expedient from different (computational) points of view \cite{roden2012}. That is, in the simulation of molecules embedded in an environment, one might explicitly include one or more strongly coupled (important) vibrational modes in the system part and treat the remaining part effectively. Alternatively, one could consider only the electronic degrees of freedom of the molecules as system part, which in turn are coupled to a highly structured environment. 
The very same considerations apply in the case of some electronic degrees of freedom being coupled to an imperfect (lossy) cavity \cite{luoma2014,shabani2014}.
While an unstructured environment is more easily handled in simulations, the rapid growth of the Hilbert space associated with the explicitly treated environmental modes makes the second choice the favorable one. 

The effect of different system-environment partitioning has already been discussed in literature \cite{garg1985,hughes2009,martinazzo2011,roden2012}, however, the discussion mostly focused on the SD. In this work, we study how the BCF transforms under different system-environment partitionings. In particular, we examine the effect of two different initial states, a factorizing and a correlated one, on the transformed BCF.

We consider an exemplary model consisting of a harmonic bath coupled to a  single mode, called ``pseudomode'' (PM). (See \rref{roden2012} for a discussion of how a PM relates to a vibrational mode.) This PM we then couple to another harmonic oscillator that acts as a system, thus allowing us to study the influence of the different BCFs on a system dynamics. For strong coupling between PM and both system oscillator and ohmic bath we find pronounced differences in the dynamics of the mean occupation number of the system oscillator, thus stressing the importance to take heed of the initial state of the composite system in this regime.

The paper is structured as follows: In Sec.~\ref{sec:analytic} we introduce the microscopic model on which our discussion is based. We outline the procedure according to which BCFs transform and state analytic formulae for the case of a single PM coupled to a harmonic bath. In Sec.~\ref{sec:numerics}, we evaluate the transformed BCFs numerically and discuss some examples, highlighting the regimes in which notable differences are induced by different initial states. Finally, we summarize our findings in Sec.~\ref{sec:conclusion}.

Details of calculations are given in three appendices: In Appendix~\ref{sec:app_SD_BCF} we review the definition of BCF and SD on basis of the microscopic model introduced in the main text. In Appendix~\ref{sec:app_bath_trafos}, we explain how the PM and bath operators can be transformed into a basis in which the combined Hamiltonian of PM and bath is diagonal. Lastly, we review in Appendix~\ref{sec:app_heisenberg_eqns} an alternative derivation of the transformed BCFs on grounds of the Heisenberg equations of motion and state a numerical recipe to solve the occurring integro-differential equation.

\section{Model system and analytic transformations}\label{sec:analytic}
In this section, we detail model Hamiltonian and framework necessary to perform the analytic transformation of the BCF presented at the end of the section.
To that end, we first review the standard model of a system linearly coupled to an environment of independent oscillators (Sec.~\ref{subsec:standard_model}) and introduce the model Hamiltonian we consider in this work (Sec.~\ref{subsec:model_H}). Subsequently, we discuss two particular ways to partition this Hamiltonian into a system and an environment part (Sec.~\ref{subsec:sys_env_partitioning}) and explain how the transformed BCF can be calculated (Sec.~\ref{subsec:BCF_calculation}). Finally, we specify two initial environment states and present the corresponding transformed BCFs (Sec.~\ref{subsec:initial_states}).

\subsection{General properties of a system linearly coupled to an environment of independent oscillators}\label{subsec:standard_model}
We start our discussion by reviewing the standard model of a system linearly coupled to an environment of independent oscillators, which will allow us later on to point out differences in the transformed BCFs most clearly.

In the standard model, the total Hamiltonian is partitioned into three parts,
\begin{equation}\label{eq:H_tot}
 \mathcal{H}_\mathrm{tot} = \mathcal{H}_\mathrm{S} + \mathcal{H}_\mathrm{S-E} + \mathcal{H}_\mathrm{E},
\end{equation}
where $\mathcal{H}_\mathrm{S}$ denotes the Hamiltonian of the system, containing the degrees of freedom in which one is interested in, and
\begin{equation}\label{eq:H_bath_diagonal_c}
 \mathcal{H}_\mathrm{E}=\sum_{\mu} \tilde{\omega}_{\mu}c^{\dagger}_{\mu}c_{\mu}
\end{equation}
the Hamiltonian of the environment. Here, $c_{\mu}$ is the annihilation operator of an environment mode with frequency $\tilde{\omega}_{\mu}$. An environment of the form of $\mathcal{H}_\mathrm{E}$ we call \emph{diagonal}, meaning that the oscillators are uncoupled. 

The Hamiltonian accounting for the interactions between system and environment reads
\begin{equation}\label{eq:H_sys_env_coupling}
 \mathcal{H}_\mathrm{S-E} = \left(L_\mathrm{S} \sum_{\mu} k_{\mu}c^\dagger_{\mu} + \mathrm{H.c.}\right).
\end{equation}
The coupling Hamiltonian $\mathcal{H}_\mathrm{S-E}$ couples the environment modes $c_\mu$ via $L_\mathrm{S}$ linearly to the system, with strength $k_{\mu}$.
It is convenient to introduce the so-called spectral density encoding this frequency-dependent coupling as ($\omega > 0$)
\begin{equation}\label{eq:SD_def_posfreq}
 J(\omega) = \sum_\mu{|k_\mu|^2\delta(\omega-\tilde{\omega}_\mu)}.
\end{equation}
Note that we set $\hbar=k_\mathrm{B}=1$ throughout this work.

The relevant quantity typically entering open quantum system approaches like Redfield \cite{weiss2008,may2011}, Caldeira-Leggett \cite{weiss2008,breuer2002} and Non-Markovian Quantum State Diffusion (NMQSD) \cite{strunz1996,diosi1998} is the bath correlation function, --- for Hermitian $L_\mathrm{S}$ --- given by
\begin{equation}
\label{eq:alpha_general}
 \alpha(t,t') = \mathrm{Tr}_\mathrm{E}\left\{\left(C(t) + C^\dagger(t)\right)\left(C(t') + C^\dagger(t')\right)\hat{\rho}_\mathrm{E}(0)\right\},
\end{equation}
with $C(t)$ defined as
\begin{equation}\label{eq:C_op_timeevol}
 C(t) = e^{i\mathcal{H}_\mathrm{E} t}\sum_\mu\left(k_\mu^* c_\mu\right)e^{-i\mathcal{H}_\mathrm{E} t} \equiv \sum_\mu k_\mu^* c_\mu(t).
\end{equation}
To obtain \eref{eq:alpha_general} in the given form, the total initial state is taken to be 
\begin{equation}\label{eq:rho_0_tot_factor}
 \hat{\rho}_\mathrm{tot}(0) = \hat{\rho}_\mathrm{S}(0)\otimes\hat{\rho}_\mathrm{E}(0).
\end{equation}
This implies that no correlations between system and environment exist before the interaction between system and environment is `turned on'. This assumption, which is typically introduced by virtue of the ease of computation, establishes the significance of system-environment partitioning.

For a non-Hermitian system operator $L_\mathrm{S}$, the BCF is no longer given by \eref{eq:alpha_general}. Rather, two correlation functions are required \cite{ritschel2015}, reading
\begin{subequations}
\begin{align}
 \alpha_1(t,t') &= \mathrm{Tr}_\mathrm{E}\left\{C(t)C^\dagger(t')\hat{\rho}_\mathrm{E}(0)\right\} \,\mathrm{and} \label{eq:2BCF_microscopical_1}\\
 \alpha_2(t,t') &= \mathrm{Tr}_\mathrm{E}\left\{C^\dagger(t)C(t')\hat{\rho}_\mathrm{E}(0)\right\}.\label{eq:2BCF_microscopical_2}
\end{align}
\end{subequations}

If the BCF is stationary, i.e., if $\alpha(t,t')$ is a function of the time difference only, $\alpha(t,t')=\alpha(t-t',0)$, it is convenient to write $\alpha(\tau)\equiv\alpha(\tau,0)$, with $\tau=t-t'$. Note that the stationarity of the BCF depends on the initial state of the environment in general.

If the initial state of the diagonal environment $\hat{\rho}_\mathrm{E}(0)$, which enters \eref{eq:alpha_general}, is a thermal state (and if $L_\mathrm{S}$ is Hermitian), the BCF of the environment is of the `standard' form
\begin{equation}\label{eq:BCF_coth_def}
 \alpha(\tau) = \int_0^\infty d\omega J(\omega)\left(\coth\left(\frac{\omega}{2T}\right)\cos(\omega \tau)-i\sin(\omega \tau)\right),
\end{equation}
with $\tau=t-t'$. In \eref{eq:BCF_coth_def}, $T$ is the temperature of the environment and $J(\omega)$ the SD.
For a detailed review of SD and BCF in the standard case, see Appendix~\ref{sec:app_SD_BCF}.

\subsection{Model Hamiltonian with PM}\label{subsec:model_H}
We now introduce the total Hamiltonian which we focus on in this work, which is of the form
\begin{equation}\label{eq:H_tot_detail}
 \mathcal{H}_\mathrm{tot} = \mathcal{H}_\mathrm{rel} + \mathcal{H}_\mathrm{rel-PM}+\mathcal{H}_\mathrm{PM} + \mathcal{H}_\mathrm{PM-B} + \mathcal{H}_\mathrm{B}.
\end{equation}
In \eref{eq:H_tot_detail}, $\mathcal{H}_\mathrm{rel}$ contains the relevant degrees of freedom we are interested in. This relevant part is via the Hamiltonian 
\begin{equation}\label{eq:H_system_PM_coupling}
  \mathcal{H}_\mathrm{rel-PM} = \left(g^* b L^\dagger + \mathrm{H.c.}\right)
\end{equation}
linearly coupled to the PM, whose Hamiltonian reads
\begin{equation}\label{eq:H_PM_bath_def}
 \mathcal{H}_\mathrm{PM} =  \Omega b^\dagger b.
\end{equation}
Here, $L$ is some operator in the Hilbert space of the relevant Hamiltonian $\mathcal{H}_\mathrm{rel}$, $g$ a coupling constant quantifying the strength of the coupling, and $\Omega$ the frequency of the PM with annihilation operator $b$. In addition, the PM is coupled to a diagonal bath
\begin{equation}\label{eq:H_B}
\mathcal{H}_\mathrm{B} = \sum_\lambda{\omega_\lambda a_\lambda^\dagger a_\lambda},
\end{equation}
where $\omega_\lambda$ are the frequencies belonging to the bath modes $\lambda$ with annihilation operators $a_\lambda$. 
The coupling Hamiltonian $\mathcal{H}_\mathrm{PM-B}$ is taken to be bilinear,
\begin{equation}\label{eq:H_PM_bath_coupling}
 \mathcal{H}_\mathrm{PM-B} = \sum_\lambda{\left(\kappa_\lambda^* a_\lambda b^\dagger + \mathrm{H.c.}\right)},
\end{equation}
with $\kappa_\lambda$ being the coupling constants quantifying the coupling between PM and bath modes.
The generalization of our discussion to several PMs is straightforward in many cases of interest (e.g., for single linear chains of PMs \cite{martinazzo2011a,woods2014} and multiple linear chains \cite{huh2014}), and will be addressed at the end of the section.

\subsection{System-environment partitioning}\label{subsec:sys_env_partitioning}
We now consider two particular examples of assigning the PM to the different parts of the total Hamiltonian, illustrated in \fref{fig:bath_illustration}.
This leads to different choices of the system Hamiltonian $\mathcal{H}_\mathrm{S}$, the environment Hamiltonian $\mathcal{H}_\mathrm{E}$, and the coupling between them. We denote the two different ways of partitioning by SI and SII for the system and EI and EII for the environment, respectively.

\subsubsection{PM in the system}\label{subsec:PM_system}
The first partitioning [(I) in \fref{fig:bath_illustration}] is to take the PM as part of the system, which amounts to setting
\begin{align}
 \mathcal{H}_\mathrm{SI} & = \mathcal{H}_\mathrm{rel} + \mathcal{H}_\mathrm{rel-PM} + \mathcal{H}_\mathrm{PM},\label{eq:H_Stilde}\\ 
 \mathcal{H}_{\mathrm{EI}} & = \mathcal{H}_{B},
\end{align}
and
\begin{equation}
 \mathcal{H}_{\mathrm{SI}-\mathrm{EI}}=\mathcal{H}_\mathrm{PM-B}.
\end{equation}
Note that the system now contains beside the relevant degrees of freedom also the PM, and that the environment is in the standard form of \eref{eq:H_tot}. 

\subsubsection{PM in the environment}\label{subsec:PM_bath}
The second partitioning is illustrated in panel (II) in \fref{fig:bath_illustration}. Here, the system is given by 
\begin{equation}
\mathcal{H}_\mathrm{SII}=\mathcal{H}_\mathrm{rel},
\end{equation}
while the environment $\mathrm{EII}$ contains both PM and bath,
\begin{equation}\label{eq:H_Btilde}
 \mathcal{H}_{\mathrm{EII}} = \mathcal{H}_\mathrm{PM} + \mathcal{H}_\mathrm{PM-B} + \mathcal{H}_\mathrm{B}.
\end{equation}
Accordingly, the coupling between system and environment is given by
\begin{equation}
 \mathcal{H}_{\mathrm{SII}-\mathrm{EII}} = \mathcal{H}_\mathrm{S-PM}.
\end{equation}
Note that the resulting Hamiltonian is \emph{not} in the standard form of \eref{eq:H_tot}, as the environment is not diagonal. It can, however, be diagonalized by a simple transformation, as detailed in Appendix~\ref{sec:app_bath_trafos}.

\subsection{Calculation of the BCF}\label{subsec:BCF_calculation}
In terms of the SD, it is known how to transform between different ways of system-environment partitioning \cite{garg1985,garraway1997,martinazzo2011,roden2011,roden2012}. For this procedure, it is sufficient to know the total Hamiltonian, as the SD is fully encoded in $\mathcal{H}_\mathrm{tot}$.

\begin{figure}[tb]
\centering
\includegraphics[width=\columnwidth]{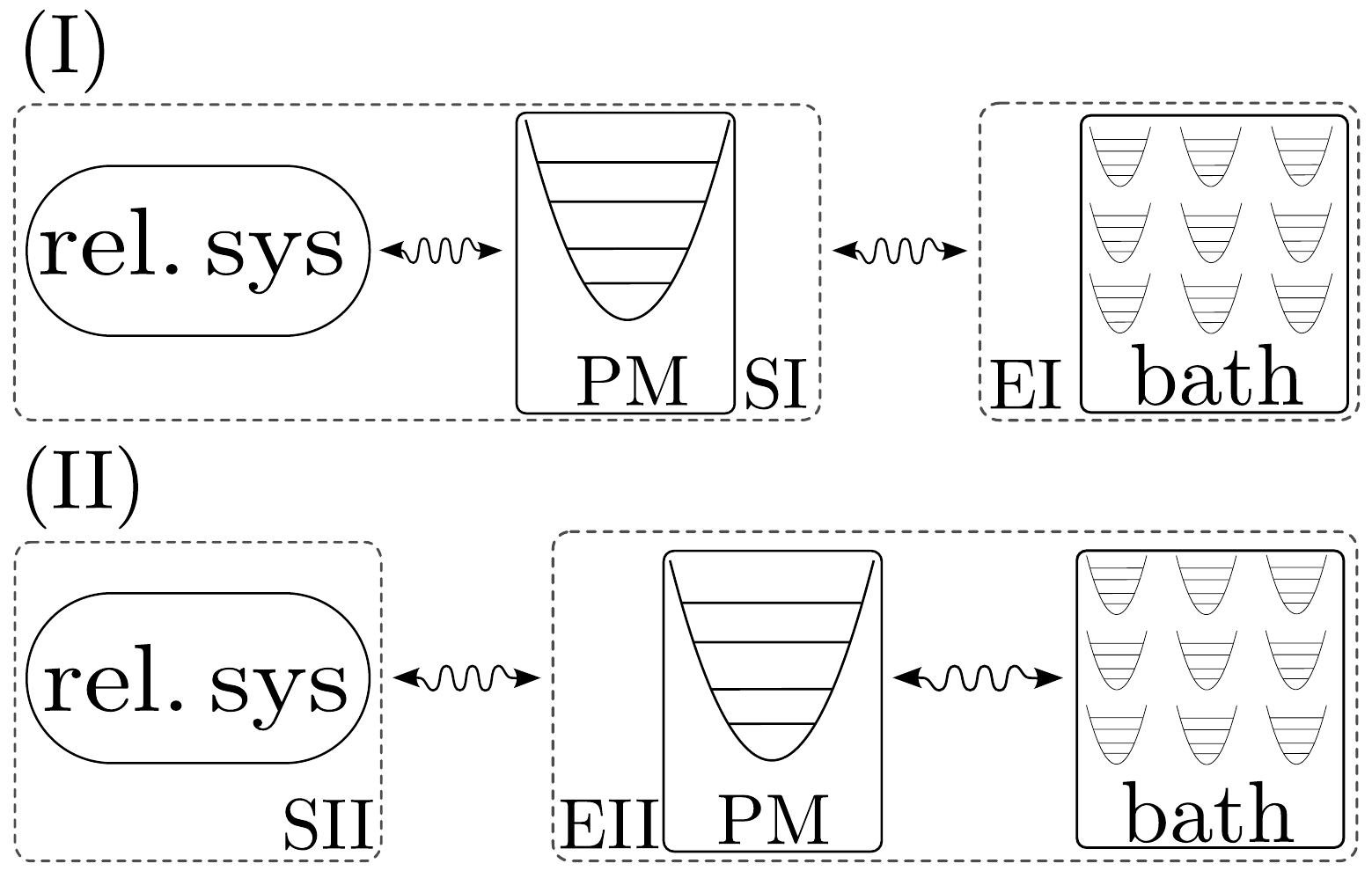}
\caption{Illustration of two different ways of performing a system-environment partitioning in the presence of a PM linearly coupled to both system and bath. In (I), the system part SI consists of the relevant system degrees of freedom linearly coupled to the PM, which in turn is coupled to an (unstructured) environment EI whereas in (II) the system SII directly couples to a (structured) environment EII including the PM.} 
\label{fig:bath_illustration}
\end{figure}

The BCF, however, depends on the environment state, denoted by $\hat{\rho}_\mathrm{EI}(0)$ and $\hat{\rho}_\mathrm{EII}(0)$ respectively for the two settings (I) and (II) in \fref{fig:bath_illustration}. 
Note that to obtain a total initial state of form \eref{eq:rho_0_tot_factor}, we need $\hat{\rho}_\mathrm{tot}(0) = \hat{\rho}_\mathrm{SI}(0)\otimes\hat{\rho}_\mathrm{EI}(0)$ in setting (I) whereas $\hat{\rho}_\mathrm{tot}(0) = \hat{\rho}_\mathrm{SII}(0)\otimes\hat{\rho}_\mathrm{EII}(0)$ in setting (II).

Microscopically, the BCF is the (two-time) correlation function of the environment operators in the system-environment coupling. In setting (I), the environment is diagonal and we can therefore directly use \eref{eq:alpha_general}. For setting (II), in contrast, the environment EII is not diagonal. Nonetheless, we can similarly to \eref{eq:C_op_timeevol} write down the time evolution of the environment coupling operator,
\begin{equation}
 B(t) = g^*\,e^{i\mathcal{H}_{\mathrm{EII}} t}\,b\,e^{-i\mathcal{H}_{\mathrm{EII}} t} \equiv g^*\, b(t),
\end{equation}
whose time dependence arises via transformation into the interaction picture with respect to $\mathcal{H}_{\mathrm{EII}}$.

For a Hermitian system operator, $L=L^\dagger$, and $\mathcal{H}_\mathrm{S-PM}$ can be written as $\mathcal{H}_\mathrm{S-PM} = L (g^* b + g b^\dagger)$, which has a Hermitian environment coupling operator. In this case, the BCF is given by
\begin{align}
 \alpha(t,t') &= \mathrm{Tr}_{\mathrm{EII}}\left\{\left(B(t) + B^\dagger(t)\right)\left(B(t') + B^\dagger(t')\right)\hat{\rho}_{\mathrm{EII}}(0)\right\}\nonumber\\
 &\equiv \left\langle\left(B(t) + B^\dagger(t)\right)\left(B(t') + B^\dagger(t')\right)\right\rangle_\mathrm{EII} \label{eq:BCF_microscopical},
\end{align}
where $\hat{\rho}_{\mathrm{EII}}(0)$ denotes the initial density operator of the environment and the subscript $\mathrm{EII}$ of the trace indicates that the trace is taken over the environmental degrees of freedom. 

To evaluate the BCF \pref{eq:BCF_microscopical}, it is convenient to take advantage of the existence of a linear transformation between the PM operator $b$ and the operators in which the Hamiltonian $\mathcal{H}_{\mathrm{EII}}$ and the initial state, respectively, are diagonal (cf.\ Appendix~\ref{sec:app_bath_trafos}). 
Specifically, we first transform the PM operator into the basis in which the environment Hamiltonian $\mathcal{H}_{\mathrm{EII}}$ is diagonal, $b(t) = [S\bar{c}(t)]_0$, by means of the transformation matrix $S$. The time evolution of the annihilation operators $c_\mu$ of the diagonal Hamiltonian $\mathcal{H}_{\mathrm{EII}}$, however, is simply given by $c_\mu(t) = e^{-i\tilde{\omega}_\mu t} c_\mu$. Subsequently, the operators $c_\mu$ are transformed into the basis in which the initial state is diagonal, if necessary, and the BCF is evaluated.

\subsection{Choice of initial states of the environment}\label{subsec:initial_states}
As discussed in the previous section, the total initial state in case (I) is typically taken to be $\hat{\rho}_\mathrm{tot}(0) = \hat{\rho}_\mathrm{SI}(0)\otimes\hat{\rho}_\mathrm{EI}(0)$. When moving the PM from the system part to the environment part, i.e., going from (I) to (II), one could thus reason that the initial state of the environment EII should be given by $\hat{\rho}_\mathrm{EII}(0) = \hat{\rho}_\mathrm{PM}(0)\otimes\hat{\rho}_\mathrm{EI}(0)$. 
Conversely, if one considers the PM to be part of the environment EII from the very beginning on, there is no reason why the PM should be uncorrelated with EI. From this point of view, a correlated initial state between PM and EI seems to be more `natural'.
To clarify the implications of the two aforementioned choices we employ these initial states when evaluating the BCF in setting (II) in the following.

\begin{figure*}[tb]
\centering
\includegraphics[width=0.8\paperwidth]{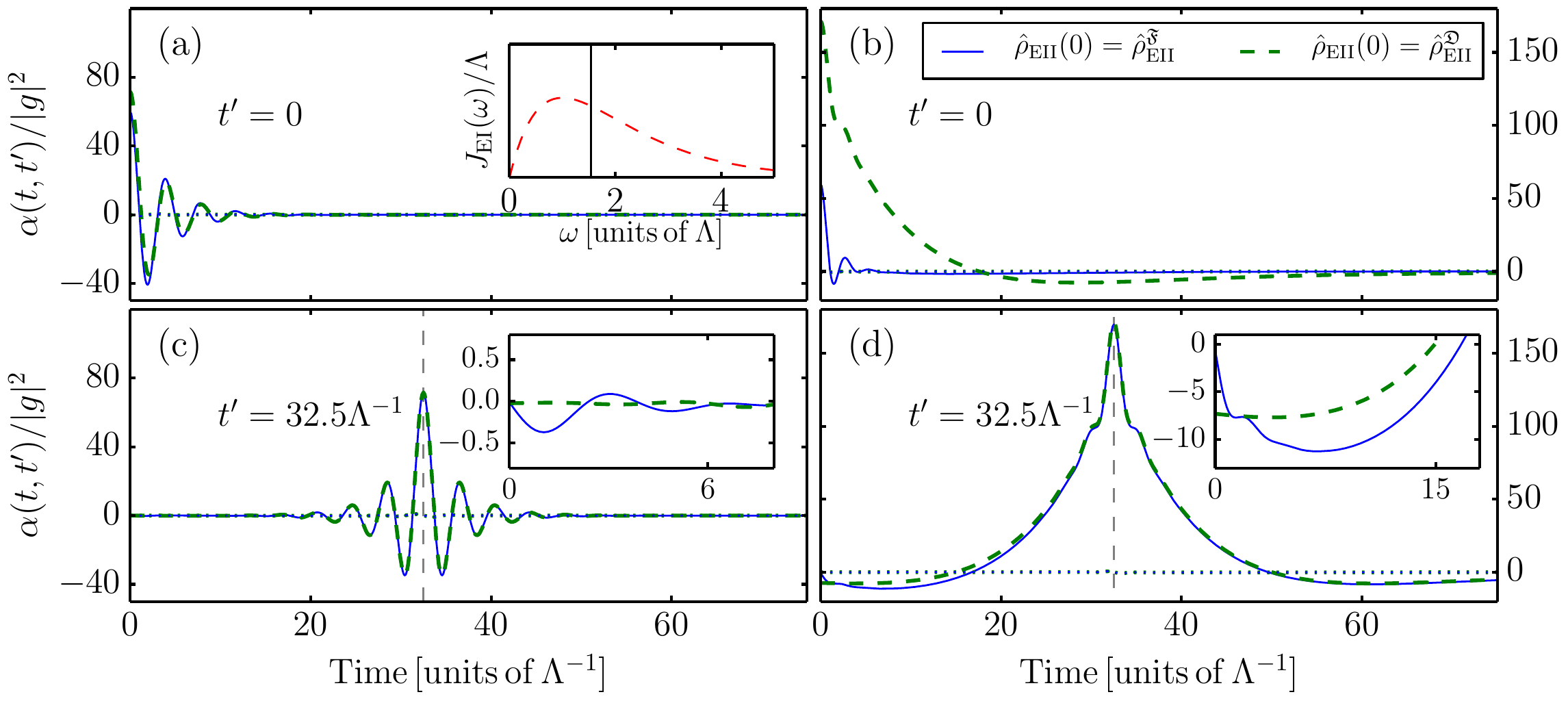}
\caption{(Color online) Bath correlation functions $\alpha(t,t')/|g|^2$ for different coupling strengths $\eta$ and different reference times $t'$. The left column [panels (a),(c)] shows the BCF for weak PM-bath coupling, $\eta=0.25$, whereas the right column [panels (b),(d)] shows a strong coupling regime, $\eta=1.0$. In the first row [panels (a)-(b)] the reference time of the BCF is set to $t'=0$, in the second row [panels (c)-(d)] $t'=32.5\,\Lambda^{-1}$. Solid blue lines indicate the real part of the BCF with factorizing initial state [\eref{eq:rho_0_factor}], dashed green lines the real part of the BCF with diagonal initial state [\eref{eq:rho_0_diag}]. Dotted lines show the corresponding imaginary parts. The insets in panel (c)-(d) provide a detail of the short-time dynamics, while the inset in (a) shows the SD $J_\mathrm{EI}(\omega)$ (dashed red line) with the position of the PM indicated by a solid vertical line. The BCFs were calculated using $T=46\,\Lambda$ and $\Omega=1.5\,\Lambda$.} 
\label{fig:numerical_example_BCF}
\end{figure*}

\subsubsection{Factorizing initial state between PM and EI}
The first initial state we consider is the one typically associated with \textit{factorizing} initial conditions between PM and bath EI,
\begin{equation}\label{eq:rho_0_factor}
 \hat{\rho}_{\mathrm{EII}}^\mathfrak{F}(0) = \frac{1}{Z} e^{-\beta\left(\Omega b^\dagger b + \sum_\lambda{\omega_\lambda a_\lambda^\dagger a_\lambda}\right)},
\end{equation}
where $\beta=1/T$ is the inverse temperature and the partition function $Z$ is defined such that $\mathrm{Tr}_{\mathrm{EII}}\{\hat{\rho}_{\mathrm{EII}}^\mathfrak{F}(0)\} = 1$. This particular initial state is widely used, owing to its convenient properties in analytic calculations. 
The physical assumption implied is that during thermal equilibration with an ambient heat bath no correlations are built up between PM and bath EI; a reasoning that only holds in the limit of vanishing coupling between PM and bath since only in this limit independent thermal equilibration (i.e., equilibration to the respective canonical states) of two coupled systems exists \cite{kubo1991,chaudhry2013,yang2014,iles-smith2014} (cf.\ also \rrefs{kampen2004,ambegaokar2007}).

Evaluating the BCF \eref{eq:BCF_microscopical} with factorizing initial state, we find (see Appendix~\ref{sec:app_bath_trafos} for details)
\begin{align}
  \alpha(t,t') &= |g|^2\sum_{\mu,\nu,\eta} \left[S_{0\mu}^* S_{0\nu} S_{\eta\mu} S_{\eta\nu}^* e^{i(\tilde{\omega}_\mu t -\tilde{\omega}_\nu t')} n(\omega_\eta) \right.\nonumber\\
  &\quad\left. +\,S_{0\mu} S_{0\nu}^* S_{\eta\mu}^* S_{\eta\nu} e^{-i(\tilde{\omega}_\mu t - \tilde{\omega}_\nu t')} \left(n(\omega_\eta)+1\right)\right].\label{eq:BCF_singlePM_factor_num}
\end{align}
Here, $n(\omega)$ is the mean occupation number of an environment oscillator with frequency $\omega$,
\begin{equation}
 n(\omega) = \frac{1}{(e^{\beta\omega}-1)}.
\end{equation}
Note that \eref{eq:BCF_singlePM_factor_num} is not in the form of \eref{eq:BCF_coth_def}; in fact, we cannot even write $\alpha(t,t') = \alpha(t-t',0)$.

\subsubsection{Thermal (correlated) state of PM and EI}
The second initial state we consider we call \textit{diagonal} initial state; we thereby denote the canonical state of the environment $\mathrm{EII}$. This state we obtain for increased coupling between PM and bath EI, since with increasing coupling the equilibrium state of PM and bath will be given by the thermal state of the joint PM-bath environment, which no longer factorizes into a PM and a bath part. Introducing the creation (annihilation) operators of the eigenmodes of the joint PM-bath system $c_\mu^\dagger$ ($c_\mu$), the global thermal state reads
\begin{equation}\label{eq:rho_0_diag}
 \hat{\rho}_{\mathrm{EII}}^\mathfrak{D}(0) = \frac{1}{Z} e^{-\beta\sum_\mu{\tilde{\omega}_\mu c_\mu^\dagger c_\mu}}.
\end{equation}
Here, the superscript $\mathfrak{D}$ denotes a diagonal initial PM-bath state, implying that at $t=0$ PM and bath have jointly evolved towards a thermal state whose occupation depends on the eigenenergies $\tilde{\omega}_\mu$ of the composite system. As before, $Z$ is defined such that $\mathrm{Tr}_{\mathrm{EII}}\{\hat{\rho}_{\mathrm{EII}}^\mathfrak{D}(0)\} = 1$.

For the diagonal initial state the BCF reads ($\tau=t-t'$)
\begin{equation}\label{eq:BCF_singlePM_diagonal_num}
  \alpha(\tau) = |g|^2\sum_\mu |S_{0\mu}|^2 \left[e^{i\tilde{\omega}_\mu\tau} n(\tilde{\omega}_\mu) +e^{-i\tilde{\omega}_\mu\tau} \left(n(\tilde{\omega}_\mu)+1\right)\right].
\end{equation}
Note that \eref{eq:BCF_singlePM_diagonal_num} is of the same form as in the standard case [cf.\ \eref{eq:standard_BCF_explicit}] and can hence be written in the standard form \eref{eq:BCF_coth_def} with transformed SD.

\subsubsection{Discussion}
Equations \pref{eq:BCF_singlePM_factor_num} and \pref{eq:BCF_singlePM_diagonal_num} allow for several observations. Firstly, the BCF comprises of the time evolution of the eigenmodes of the PM-bath environment weighted by the populations of the eigenmodes in the initial state.
Secondly, we explicitly see that in case of a diagonal initial state we obtain a stationary BCF, whereas in the case of factorizing initial conditions the BCF is non-stationary (for small times). This is to be expected, since for a PM-bath environment in thermal equilibrium the expectation value of any number operator (e.g.\  $b^\dagger b$) should not depend on time --- which is exactly what we observe if we set $\tau=0$ in \eref{eq:BCF_singlePM_diagonal_num}. (As $\alpha(t,t')\propto \langle b^\dagger(t)b(t') + b(t)b^\dagger(t')\rangle_\mathrm{EII}$, the stationarity of $\langle b^\dagger(t) b(t) \rangle_\mathrm{EII}$ can be directly read off from the BCF.)

The procedure outlined above can be generalized straightforwardly to, e.g., linearly-coupled chains of PMs of which the last PM is possibly coupled to a diagonal harmonic bath \cite{martinazzo2011,woods2014}, directly coupled PMs with independent baths \cite{ritschel2014}, or a combination of both \cite{huh2014}.
As the BCF of the system is determined by the correlation function of the PM operator directly coupled to the system, we simply need to adjust $\mathcal{H}_{\mathrm{EII}}$ in the above treatment; the calculation of the BCF then proceeds in the exact same manner as detailed above.

Since the neglect of initial correlations can lead to noticeable differences in the dynamics \cite{hakim1985,pollak2008,chaudhry2013}, depending on the parameters of the underlying Hamiltonian, we now turn to the discussion of numerical examples.

\section{Numerical examples}\label{sec:numerics}

\subsection{Evaluation of the transformed BCFs}\label{subsec:num_eval_BCF}
In our numerical examples, we take as spectral density for EI an ohmic SD with exponential cutoff ($\omega > 0$),
\begin{equation}\label{eq:SD_ohmic_cutoff}
 J_\mathrm{EI}(\omega) = \eta \omega e^{-\omega/\Lambda},
\end{equation}
where $\Lambda$ is the cutoff frequency and $\eta$ a scaling for the overall coupling strength. For numerical purposes, we sample $J_\mathrm{EI}(\omega)$ at discrete frequencies $\omega_\lambda$. The couplings  $\kappa_\lambda$ of \eref{eq:H_PM_bath_coupling} we obtain by evaluating the quadrature \cite{roden2012}
\begin{equation}
 \kappa_\lambda = \sqrt{J(\omega_\lambda)\Delta\omega_\lambda},
\end{equation}
with $\Delta\omega_\lambda = (\omega_{\lambda+1}-\omega_{\lambda-1})/2$ for $\lambda=2,\dotsc,N-1$; $\Delta\omega_1=\omega_2-\omega_1$ and $\Delta\omega_N=\omega_N-\omega_{N-1}$.

\begin{figure*}[tb]
\centering
\includegraphics[width=0.8\paperwidth]{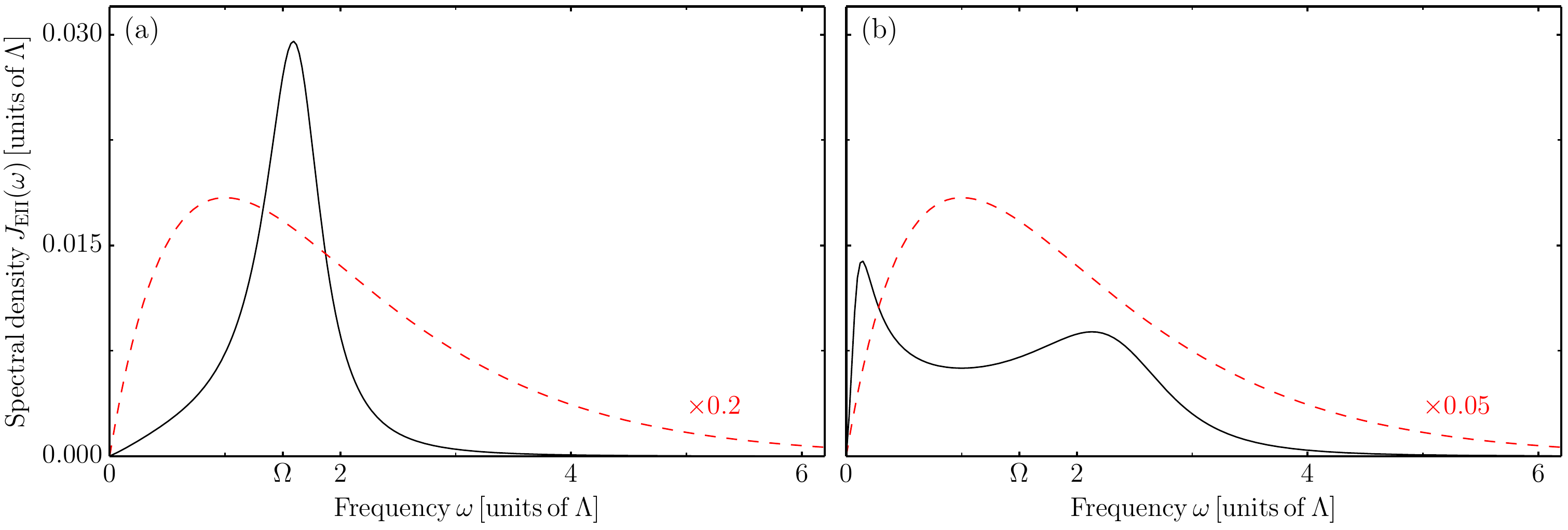}
\caption{(Color online) Spectral density $J_\mathrm{EII}(\omega)$ (black, solid line) for (a) $\eta=0.25$ and (b) $\eta=1.0$ obtained from Fourier transformation of the BCF (for details, see text). As the BCFs for diagonal and factorizing initial state (evaluated at $t_\mathrm{cm} = 130\,\Lambda^{-1}$) coincide, they yield identical SDs. In both panels, the red, dashed line indicates the original SD of the ohmic bath, which has been scaled by the factor denoted in red. All other parameters are as in \fref{fig:numerical_example_BCF}.} 
\label{fig:numerical_example_SD}
\end{figure*}

The sampling range is chosen such that the full SD is covered. For the particular cases shown, we use $N=4000$ bath oscillators for the numerical discretization, with equidistantly spaced frequencies, starting from $0.002\,\Lambda$.
For the sake of clarity of presentation, we choose the PM frequency close to the maximum of the ohmic SD, setting $\Omega=1.5\,\Lambda$. This choice renders the coupling between PM and bath strongly dependent on the overall scaling of the SD, which is quantified by $\eta$.

Note that the Hamiltonian $\mathcal{H}_\mathrm{EII}$ is positive for all parameters employed in the numerical calculations discussed in this section, and that the finite recurrence time of the BCF is large enough to observe complete decay of the BCF.

We now evaluate the BCFs \esref{eq:BCF_singlePM_factor_num}{eq:BCF_singlePM_diagonal_num} with the described numerical procedure for different couplings $\eta$ and times $t'$. Using the SD of \eref{eq:SD_ohmic_cutoff}, this yields the BCFs displayed in \fref{fig:numerical_example_BCF}.
There, the left column [panels (a),(c)] corresponds to relatively weak coupling $\eta=0.25$ whereas in the right column [panels (b),(d)] the PM is relatively strongly coupled to the bath EI, $\eta=1.0$. Furthermore, the first row [panels (a),(b)] show the BCF evaluated at $t'=0$ while the second row [panels (c),(d)] show the BCF evaluated at $t'=32.5\,\Lambda^{-1}$.

As can be seen from \frefp{fig:numerical_example_BCF}{(a) and (b)}, at $t'=0$ pronounced differences emerge between the two different initial conditions (blue vs. green) as the coupling $\eta$ is increased. On the one hand, the damping of the BCF is increased in the presence of strong coupling, $\eta=1.0$ (note that the overall time of equilibration of the BCF increases as well), which results in different dynamics for the two different initial states. On the other hand, the initial values of the BCFs, $\alpha(0,0)$, change, highlighting the increasing differences between the initial states that manifest themselves in the dynamics of the BCF.

Considering the BCFs evaluated at $t'=32.5\,\Lambda^{-1}$, shown in \frefp{fig:numerical_example_BCF}{(c) and (d)}, we observe that the BCFs obtained from different initial states look very similar, with the strongest difference being a transient equilibration dynamics present for factorizing initial conditions, which gets more noticeable in the strong coupling case. The differences found between the different initial states for $t=t'=0$ have almost vanished for $t=t'=32.5\,\Lambda^{-1}$ [cf.\ panels (a, c) and (b, d), respectively], due to the fact that equilibration has already taken place before $t=32.5\,\Lambda^{-1}$.

The non-stationarity of the BCF for factorizing initial conditions is related to what is called ``initial slippage'' if a system is coupled to a Markovian environment. In such systems, a non-Markovian feature can be present at small times due to the fact that Markovian dynamics for the total system requires correlations between system and environment that are not present initially if factorizing initial conditions are employed \cite{suarez1992,weiss2008}. Hence, slippage of initial conditions can remedy non-Markovian dynamics introduced by an initially uncorrelated system-environment state. In the same manner, the BCFs displayed in \fref{fig:numerical_example_BCF} need some time to equilibrate for factorizing initial conditions before reaching `stationarity' \cite{hakim1985}.

\subsection{Corresponding SDs}\label{subsec:num_SDs}
The SD, as defined in \eref{eq:SD_def_posfreq}, is fully determined by the total Hamiltonian. Consequently, it should not depend on the initial state of the environment. For a diagonal environment, however, the SD can be extracted from the BCF owing to the relation \pref{eq:BCF_coth_def} between BCF and SD.

That is, for the diagonal initial state, the SD can be obtained by Fourier transforming \eref{eq:BCF_singlePM_diagonal_num} with respect to the time difference $\tau$ and dividing by the factor $2\pi(n(\omega)+1)$ [cf.\ \eref{eq:BCF_j}]. For factorizing initial state, we can rewrite \eref{eq:BCF_singlePM_factor_num} as a function of the center of mass coordinate $t_\mathrm{cm}=(t+t')/2$ and the time difference $\tau=t-t'$ and perform a Fourier transformation with respect to $\tau$.

The spectral densities corresponding to the parameters of \fref{fig:numerical_example_BCF} are shown in \fref{fig:numerical_example_SD}.
For large times $t_\mathrm{cm}$, $t_\mathrm{cm} \gtrsim 120\,\Lambda^{-1}$, the difference between the Fourier transformations obtained from the BCFs using a diagonal and a factorizing initial state respectively, vanish. For this reason, we show only a single SD in \fref{fig:numerical_example_SD}. 

The SDs obtained from the BCFs perfectly agree with the analytical result of \rref{roden2012}, which has been obtained by direct transformation of the SD. (Note that in \rref{roden2012} another convention for the SD has been used, i.e., the SD is defined as $\omega^2 J(\omega)$ in our convention. For a discussion of the advantage of the convention employed in this paper, see \rref{ritschel2014}.)

As shown in \fref{fig:numerical_example_SD}, for weak coupling $\eta=0.25$ the SD exhibits only a single peak at approximately the PM frequency $\Omega$, with a width  proportional to the coupling $\eta$. For large coupling $\eta=1.0$, the single peak is split into two and the coupling strength at the PM frequency is reduced. This illustrates that for large PM-bath coupling, the bath properties indeed become essential for a system coupled to the PM.

\subsection{Corresponding system dynamics}\label{subsec:num_sys_dyn}
We now turn to analyzing the effect of the features seen in \fref{fig:numerical_example_BCF} on a system observable. To that end we specify the system operator $L$ ($L^\dagger$) in \eref{eq:H_system_PM_coupling} as the annihilation (creation) operator $d$ ($d^\dagger$) of a harmonic oscillator with frequency $\Omega_\mathrm{sys}$, with the associated system Hamiltonian reading
\begin{equation}
 \mathcal{H}_\mathrm{SII} = \Omega_\mathrm{sys} d^\dagger d.
\end{equation}
This particular set of non-Hermitian coupling operators allows us to conveniently evaluate the dynamics of the total system via diagonalization, as outlined in Sec.~\ref{subsec:BCF_calculation}.
The corresponding correlation functions $\alpha_1(t,t')$ and $\alpha_2(t,t')$ defined in \esref{eq:2BCF_microscopical_1}{eq:2BCF_microscopical_2} can be easily read off from \esref{eq:BCF_singlePM_factor_num}{eq:BCF_singlePM_diagonal_num}.

\begin{figure}[tb]
\centering
\includegraphics[width=\columnwidth]{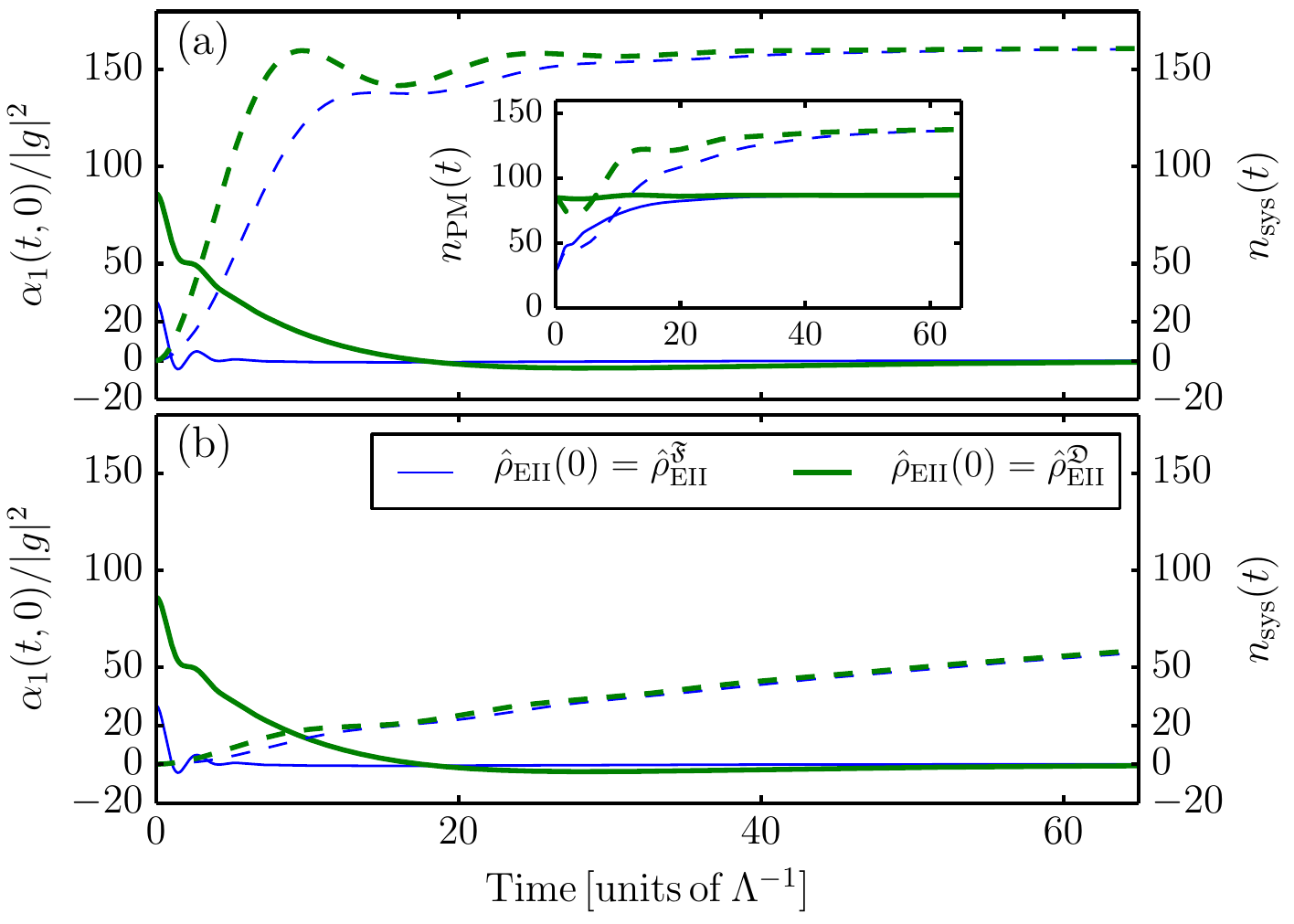}
\caption{(Color online) Bath correlation function $\alpha_1(t,0)/|g|^2$ (solid lines) and system mean occupation number $n_\mathrm{sys}(t)\equiv \langle d^\dagger(t)d(t)\rangle_\mathrm{EII}$ (dashed lines) for different initial conditions and different coupling strengths $g$. In the upper panel (a), $g=0.3\,\Lambda$, whereas in the lower panel (b), $g=0.08\,\Lambda$. Thin blue lines indicate the BCF with factorizing initial state [\eref{eq:rho_0_factor}], thick green lines indicate the BCF with diagonal initial state [\eref{eq:rho_0_diag}]. Only the real parts of the BCFs are shown. The inset shows the PM mean occupation number $n_\mathrm{PM}(t)\equiv \langle b^\dagger(t)b(t)\rangle_\mathrm{EII}$ for $g=0.3\,\Lambda$ (dashed line) and $g=0.08\,\Lambda$ (solid line) for both initial states.
Except for $\eta=1.0$ and $\Omega_\mathrm{sys} = 0.46\,\Lambda$, the parameters of \fref{fig:numerical_example_BCF} have been used. Note that $\alpha_1(t,0)$ is approximately $\alpha(t,0)/2$ in \fref{fig:numerical_example_BCF}.} 
\label{fig:numerical_example_nsys}
\end{figure}

Setting in the above parameters $\eta=1.0$, $\Omega_\mathrm{sys} = 0.46\,\Lambda$, and requiring $n_\mathrm{sys}(0) \equiv \langle d^\dagger(0)d(0)\rangle_\mathrm{EII} = 0$, we obtain \fref{fig:numerical_example_nsys} for two different system-PM couplings, $g/\Lambda=0.3$ and $g/\Lambda=0.08$. Note that only $\alpha_1(t,t')$ is shown in \fref{fig:numerical_example_nsys} since for the parameters chosen, $\alpha_2(t,t')$ is indistinguishable from  $\alpha_1(t,t')$ on the scale of the figure. The reason for choosing $\Omega_\mathrm{sys}$ relatively small is that the steady state value of $n_\mathrm{sys}(t)$ decreases with increasing $\Omega_\mathrm{sys}$, such that absolute differences in $n_\mathrm{sys}(t)$ are suppressed for large system frequencies. Likewise, we have to choose $\eta$ large in order to assure that the system dynamics is affected by the bath, being mediated via the PM.

For strong coupling of the PM to both system [$g=0.3\,\Lambda$, \frefp{fig:numerical_example_nsys}{(a)}] and bath, we observe a marked difference in the mean occupation number $n_\mathrm{sys}(t)$ between the results attained from using different initial states. This difference highlights that in case of a strongly coupled PM the initial state of the environment becomes important for the transient system dynamics. In contrast, the equilibrium values of both $n_\mathrm{sys}(t)$ and $n_\mathrm{PM}(t)$ are independent of the initial state \cite{hakim1985}. If we decrease the coupling $g$ to the system mode (or, similarly, the coupling $\eta$ to the bath oscillators), the difference in the dynamics decreases [cf.\ \frefp{fig:numerical_example_nsys}{(b)}] and the appropriate choice of initial conditions of the environment becomes less important. The same applies for choosing $\Omega_\mathrm{sys}$ large, as this results in a lower steady-state value of  $n_\mathrm{sys}(t)$ and consequently smaller overall deviations.

Besides, the time it takes for the system to equilibrate increases as $g$ is decreased [cf.\ \frefp{fig:numerical_example_nsys}{(b)}], as the equilibration of the system only proceeds via interaction with the PM.
Conversely, the PM equilibrates faster for small system-PM coupling $g$ (cf.\ inset), since in this case the strong PM-bath coupling dominates the equilibration dynamics of the PM.
This again illustrates that for small system-PM coupling, it is indeed valid to assume a factorized initial environment state since the differences induced by the BCF rapidly vanish from the system's point of view.

For low bath temperature, the differences in the dynamics persist, however, they become hardly observable due to the steady-state values (as well as the initial conditions) being significantly smaller as compared to high bath temperature. Hence, at low temperature, absolute deviations are reduced while relative deviations are preserved.

Our numerical simulations show for different initial environment states pronounced differences in the transient dynamics of a system that via a PM strongly couples to an ohmic bath. Thus, any scheme sensitive to the transient behavior of the BCF crucially depends on the choice of initial conditions of the total system.

\section{Conclusions}\label{sec:conclusion}
We have analytically and numerically studied the the BCF resulting from effectively treating a harmonic oscillator (PM) linearly coupled to a harmonic bath as part of the bath, for a factorizing and a correlated initial state between PM and bath respectively. We outlined the procedure to analytically derive the transformed BCF and discussed concrete examples for regimes in which the differences in the BCFs arising from different initial states manifest themselves in a different dynamics of the system, which we take to be a harmonic oscillator linearly coupled to the PM. This establishes a simple framework to evaluate transformed BCFs of PMs coupled to a harmonic bath.

In particular, we find that in the case of a correlated (diagonal) initial state the BCF features all the properties typically assumed for a BCF (i.e., stationarity, detailed balance and the relation \eref{eq:BCF_j} between the Fourier transform of the BCF and the SD). By contrast, for a factorizing initial state these properties do not apply, owed to the non-stationarity of the BCF in this case. Only after a transient equilibration dynamics that can induce different system dynamics, they are recovered.

The significance of the BCF lies in the fact that it quantifies the effect of the environment --- including environment correlations and temperature --- on the system degrees of freedom. Hence, our analysis complements the investigation of bath transformations (e.g.\ the mapping of a structured bath consisting of many harmonic oscillators to a linear chain of oscillators \cite{martinazzo2011a,woods2014,huh2014}) which focused on the SD.

Our findings highlight that (i) the differences between the BCFs for the different initial states chosen (factorizing and diagonal initial state) are negligible for small PM-bath coupling or PM-system coupling respectively, as expected, yet (ii) these differences can have strong impact on the system dynamics if the PM is strongly coupled to both system and bath. Therefore, in case of a strongly coupled PM, an appropriate initial state of the environment has to be used for the transformation of the BCF when considering a system's dynamics in the presence of finite temperature. 
This emphasizes the relevance of accounting for correlations in strongly coupled systems, which is not specific to our particular system \cite{breuer2002,morozov2012,iles-smith2014}. 
The question which initial state is to be considered as appropriate cannot be answered a priori, but has to be answered in consideration of the specific case.

\acknowledgments
We thank Gerhard Ritschel for useful discussions.

\appendix
\section{Microscopical definitions of BCF and SD}\label{sec:app_SD_BCF}
We consider the same total Hamiltonian as introduced in \eref{eq:H_tot},
\begin{equation}
 \mathcal{H}_\mathrm{tot} = \mathcal{H}_\mathrm{S} + \mathcal{H}_\mathrm{S-E} + \mathcal{H}_\mathrm{E},
\end{equation}
consisting of a system part $\mathcal{H}_\mathrm{S}$, an environment part $\mathcal{H}_\mathrm{E}$ [\eref{eq:H_bath_diagonal_c}], and a coupling part $\mathcal{H}_\mathrm{S-E}$ [\eref{eq:H_sys_env_coupling}] accounting for the interaction between system and environment.

Following Sec.~\ref{subsec:standard_model}, the BCF for a Hermitian system coupling operator $L_\mathrm{S}$ is given by [cf.\ \eref{eq:alpha_general}]
\begin{align}
 \alpha(t,t') & = \left\langle\sum_{\mu,\mu'}\left(k_\mu^* c_\mu(t)+\mathrm{H.c.}\right)\left(k_{\mu'}^* c_{\mu'}(t')+\mathrm{H.c.}\right)\right\rangle_\mathrm{E}\nonumber\\
  & =  \sum_\mu |k_\mu|^2\left(\left\langle c_\mu(t)c_\mu^\dagger(t')\right\rangle_\mathrm{E} + \left\langle c_\mu^\dagger(t)c_\mu(t')\right\rangle_\mathrm{E}\right), \label{eq:bcf_expanded}
\end{align}
where the second equality sign holds if the initial environment density operator $\hat{\rho}_\mathrm{E}(0)$ is taken to be a function of the number operators $c_\mu^\dagger c_\mu$. Here, the angle brackets denote the trace over the environment, $\langle\cdot\rangle_\mathrm{E} = \mathrm{Tr}_\mathrm{E}\{\cdot\hat{\rho}_\mathrm{E}(0)\}$.

\subsection*{Thermal environment}
In the following, we take the environment to be in a thermal state,
\begin{equation}\label{eq:rho_0_thermal}
 \hat{\rho}_\mathrm{E}(0) = \frac{1}{Z} e^{-\beta\sum_\mu{\tilde{\omega}_\mu c_\mu^\dagger c_\mu}},
\end{equation}
where $Z$ is defined such that $\mathrm{Tr}_\mathrm{E}\{\hat{\rho}_\mathrm{E}(0)\} = 1$ and $\beta$ is the inverse temperature $\beta=1/T$. 

The time evolution of the environment operators $c_\mu$ can be calculated by means of the Heisenberg equations of motion (cf.\ Appendix~\ref{sec:app_heisenberg_eqns}),
\begin{equation}
 \partial_t c_\mu(t) = i e^{i\mathcal{H}_\mathrm{E} t} [\mathcal{H}_\mathrm{E},c_\mu]e^{-i\mathcal{H}_\mathrm{E} t} = -i \tilde{\omega}_\mu c_\mu(t),
\end{equation}
where the time dependence refers to the interaction picture with respect to $\mathcal{H}_\mathrm{E}$, $c_\mu(t)=e^{i\mathcal{H}_\mathrm{E} t}\,c_\mu\,e^{-i\mathcal{H}_\mathrm{E} t}$.

Using the usual commutation relation $[c_\mu,c_{\mu'}^\dagger]=\delta_{\mu\mu'}$ and evaluating the trace in \eref{eq:bcf_expanded} in the number basis $\ket{n_\mu}$, we find
\begin{align}
 \alpha(t,t') &= \sum_\mu |k_\mu|^2 \left(e^{-i\tilde{\omega}_\mu (t-t')}(n(\tilde{\omega}_\mu)+1) \right.\nonumber\\
  &\left.\quad +\, e^{i\tilde{\omega}_\mu (t-t')}n(\tilde{\omega}_\mu)\right),\label{eq:standard_BCF_explicit}
\end{align}
with the mean occupation number $n(\omega)$ of the environment oscillator with frequency $\omega$ defined as
\begin{equation}\label{eq:mean_occupation_number}
 n(\omega) \equiv \frac{1}{(e^{\beta\omega}-1)} = \frac{\sum_n n\,e^{-\beta\omega n}}{\sum_n{e^{-\beta\omega n}}}.
\end{equation}

Performing a Fourier transformation with respect to $\tau\equiv t-t'$, we get
\begin{align}
 \alpha(\omega) &= 2\pi \sum_\mu |k_\mu|^2 \left[\left(n(\tilde{\omega}_\mu)+1\right)\delta(\omega-\tilde{\omega}_\mu)\right.\nonumber\\
  &\left.\quad +\,n(\tilde{\omega}_\mu)\delta(\omega+\tilde{\omega}_\mu)\right].\label{eq:BCF_deltafunctions}
\end{align}

By means of the relation $n(-\omega) = -[n(\omega)+1]$ and the definition 
\begin{equation}
 j(\omega) = \sum_\mu |k_\mu|^2\delta(\omega-\tilde{\omega}_\mu),
\end{equation}
we can rewrite \eref{eq:BCF_deltafunctions}, reading 
\begin{equation}\label{eq:BCF_j}
 \alpha(\omega) = 2\pi \left(n(\omega)+1\right)\left[j(\omega)-j(-\omega)\right].
\end{equation}
Following \rrefs{valleau2012,ritschel2014}, we now define the spectral density $J(\omega)$ as
\begin{equation}\label{eq:SD_def}
 J(\omega) = j(\omega) - j(-\omega)
\end{equation}
and obtain, after performing the inverse Fourier transform and rearranging using $1+2n(\omega)=\coth[\omega/(2T)]$, the standard expression  [cf.\ \eref{eq:BCF_coth_def}]
\begin{align}
 \alpha(\tau) & = \int_{-\infty}^\infty \frac{d\omega}{2\pi}\,e^{-i\omega \tau}\alpha(\omega) \nonumber\\
  &= \int_{-\infty}^\infty d\omega\,J(\omega)\left[n(\omega)+1\right]e^{-i\omega \tau}\nonumber\\
  &= \int_0^\infty d\omega J(\omega)\left(\coth\left(\frac{\omega}{2T}\right)\cos(\omega \tau)-i\sin(\omega \tau)\right).\label{eq:BCF_coth}
\end{align}
This is the well-known result for a linearly coupled harmonic environment in thermal equilibrium.

\section{Environment transformation}\label{sec:app_bath_trafos}
Here we show how the environment Hamiltonian $\mathcal{H}_\mathrm{EII}$ comprising both PM and bath is diagonalized. That is, following \rref{roden2012}, we rewrite \eref{eq:H_Btilde} as
\begin{equation}\label{eq:H_bath_nondiagonal_matrix}
 \mathcal{H}_{\mathrm{EII}} = \bar{a}^\dagger M \bar{a},
\end{equation}
where the vector $\bar{a}$ contains all environment annihilation operators,
\begin{equation}
 \bar{a} = (b,a_1,a_2,\dotsc,a_N)^T,
\end{equation}
and the matrix $M$ all environment couplings and energies,
\begin{equation}\label{eq:coupling_matrix}
M = 
 \begin{pmatrix}
  \Omega & \kappa_1^* & \kappa_2^* & \cdots & \kappa_N^*\\
  \kappa_1 & \omega_1 & 0 & \cdots & 0\\
  \kappa_2 & 0 & \omega_2 & \ddots & \vdots \\
  \vdots & \vdots & \ddots & \ddots & 0\\
  \kappa_N & 0 & \cdots & 0 & \omega_N
 \end{pmatrix}.
\end{equation}
The Hermitian matrix $M$ can be diagonalized by means of a unitary transformation,
\begin{equation}
 M = S D S^\dagger,
\end{equation}
where the diagonal matrix $D$ contains the eigenenergies of the composite bath,
\begin{equation}
 D = 
\begin{pmatrix}
  \tilde{\omega}_0 & & 0\\
   & \ddots & \\
   0 &  & \tilde{\omega}_N
 \end{pmatrix}.
\end{equation}
With these definitions, the new annihilation operators of the environment become $\bar{c} = S^\dagger\bar{a}$ where $\bar{c} = (c_0,c_1,\dotsc,c_N)^T$. The initial 
creation and annihilation operators are obtained from the new ones via the inverse transformation $\bar{a} = S \bar{c}$. 
Note that for a discrete number $N$ of $a_\lambda$ operators there are $N+1$ $c_\mu$ operators.

\section{Alternative derivation of the BCF on basis of Heisenberg equations of motion}\label{sec:app_heisenberg_eqns}
In this section we review a method alternative to the one introduced in Sec.~\ref{subsec:BCF_calculation} to calculate the BCF for the two initial states introduced in Sec.~\ref{subsec:initial_states} [\esref{eq:rho_0_factor}{eq:rho_0_diag}] by calculating the time dependence of the PM operator $b$ not via diagonalization, but rather by means of the Heisenberg equations of motion.

The Heisenberg equations of motion for the PM operator $b$ can be easily derived by evaluating the time derivative
\begin{align}
 \partial_t b(t) & = i e^{i\mathcal{H}_{\mathrm{EII}} t} [\mathcal{H}_{\mathrm{EII}},b]e^{-i\mathcal{H}_{\mathrm{EII}} t}\nonumber\\
  & = -i \Omega b(t) - i\sum_\lambda{\kappa_\lambda^* a_\lambda(t)}\label{eq:b_ode},
\end{align}
where the time dependence as before refers to the interaction picture with respect to $\mathcal{H}_{\mathrm{EII}}$, $b(t)=e^{i\mathcal{H}_{\mathrm{EII}} t}\,b\,e^{-i\mathcal{H}_{\mathrm{EII}} t}$, and $b(0)=b$. Likewise, for $a_\lambda$ we have
\begin{align}
 \partial_t a_\lambda(t) & = i e^{i\mathcal{H}_{\mathrm{EII}} t} [\mathcal{H}_{\mathrm{EII}},a_\lambda]e^{-i\mathcal{H}_{\mathrm{EII}} t}\nonumber\\
  & = -i \omega_\lambda a_\lambda(t) - i\kappa_\lambda b(t)\label{eq:a_ode}.
\end{align}
Formally integrating \eref{eq:a_ode}, we find
\begin{equation}\label{eq:a_ode_sol}
 a_\lambda(t) = e^{-i\omega_\lambda t}a_\lambda(0) - i\kappa_\lambda \int_{0}^t ds\, e^{-i\omega_\lambda (t-s)}b(s),
\end{equation}
which we can insert into \eref{eq:b_ode}, yielding
\begin{align}
 \partial_t b(t) &= -i \Omega b(t) - \int_{0}^t ds\sum_\lambda{|\kappa_\lambda|^2 e^{-i\omega_\lambda (t-s)}b(s)}\nonumber\\
 &\quad - i\sum_\lambda \kappa_\lambda^* e^{-i\omega_\lambda t}a_\lambda(0).\label{eq:b_ode_full}
\end{align}
Defining  $K(t-s)\equiv \sum_\lambda |\kappa_\lambda|^2 e^{-i\omega_\lambda(t-s)}$, we can write the solution of \eref{eq:b_ode_full} as
\begin{equation}\label{eq:b_ode_sol_nonMarkov}
 b(t) = U(t)b(0) - i\sum_\lambda{\kappa_\lambda^*a_\lambda(0)} \int_{0}^t ds\,U(t-s)e^{-i\omega_\lambda s}, 
\end{equation}
where $U(t)$ is determined from the integro-differential equation \cite{tan2011} 
\begin{equation}\label{eq:U_propagator_ode_app}
 \partial_tU(t) = -i \Omega\,U(t) - \int_{0}^t ds\,K(t-s)U(s).
\end{equation}
Note that $U(t)$ is only defined for $t\geq 0$, with the initial condition reading $U(0)=1$.
Albeit for a continuous bath spectrum an analytic calculation of $U(t)$ for specific spectral densities is possible via Laplace transforms 
\cite{zhang2012}, we focus on a numerical scheme for solving \eref{eq:U_propagator_ode_app} in the following, relying on numerical diagonalization.

To this end, we first define auxiliary coefficients $U_\lambda(t)$ whose dependence on $t$ is via a simple exponential and the integration boundary only,
\begin{equation}\label{eq:U_lambda_def}
 U_\lambda(t) = \kappa_\lambda \int_{0}^t ds\, e^{-i\omega_\lambda (t-s)}U(s).
\end{equation}
By means of this definition, we are able to cast \eref{eq:U_propagator_ode_app} into a set of coupled equations,
\begin{subequations}
\begin{align}
 \partial_t U(t) &= -i \Omega U(t) - \sum_\lambda \kappa_\lambda^* U_\lambda(t),\label{eq:U_ODE_plus_lambda}\\
 \partial_t U_\lambda(t) &= \kappa_\lambda U(t) - i\omega_\lambda U_\lambda(t).\label{eq:U_lambda_ODE}
\end{align}
\end{subequations}
Introducing the vector $\bar{u}(t) = (U(t),U_1(t),\dotsc,U_N(t))^T$, we can rewrite \esref{eq:U_ODE_plus_lambda}{eq:U_lambda_ODE} as
\begin{equation}\label{eq:U_ODE_vectorform}
 \partial_t \bar{u}(t) = -i G\bar{u}(t),
\end{equation}
with the matrix $G$ given by
\begin{equation}\label{eq:U_ODE_matrix_definition}
 G = 
 \begin{pmatrix}
  \Omega & -i\kappa_1^* & -i\kappa_2^* & \cdots & -i\kappa_N^* \\
  i \kappa_1 & \omega_1 & 0 & \cdots & 0 \\
  i \kappa_2 & 0 & \omega_2 & \ddots & \vdots \\
  \vdots & \vdots & \ddots & \ddots & 0 \\
  i \kappa_N & 0 & \cdots & 0 & \omega_N
 \end{pmatrix}.
\end{equation}
The Hermitian matrix $G$ can be diagonalized via the transformation $T^\dagger\, G\, T = D$, yielding the eigenvalues $\bar{\omega}_\mu$. 
Defining $\bar{u} = T \bar{v}$, the differential equation \eref{eq:U_ODE_vectorform} becomes 
\begin{equation}\label{eq:Uv_ode_diagonal}
 \partial_t\bar{v}(t) = -i D \bar{v}(t),
\end{equation}
which has the simple solution $\bar{v}(t) = e^{-i D t}\bar{v}(0)$ with $\bar{v}(0) = T^\dagger \bar{u}(0)$, where $\bar{u}(0)=(1,0,\dotsc,0)^T$.
Taking the first component of the full expression $\bar{u}(t)=T  e^{-i D t} T^\dagger\bar{u}(0)$, which gives $U(t)$, and writing $[\bar{u}(0)]_\mu=\delta_{0\mu}$, we arrive at the simple expression
\begin{equation}\label{eq:U_ODE_solution_simple}
 U(t) = \sum_\mu |T_{0\mu}|^2 e^{-i\bar{\omega}_\mu t}.
\end{equation}

The above reformulation allows one to map the solution of an integro-differential equation onto an eigenvalue problem which involves only matrix multiplication and diagonalization, which is numerically more robust with respect to the numerical time step than straightforward numerical integration of \eref{eq:U_propagator_ode_app}. Note that the approach used requires $K(t-s)$ to be given as a sum of exponentials that reproduce when being differentiated with respect to time.

We can now employ the above results to obtain alternative analytic expressions for the BCFs of the two initial states introduced in \esref{eq:rho_0_factor}{eq:rho_0_diag}. 

For factorizing initial conditions, \eref{eq:rho_0_factor}, the trace in \eref{eq:BCF_microscopical} is readily evaluated, yielding
\begin{align}
 \alpha(t,t')/|g|^2 &= U(t)U^*(t')\left(n(\Omega)+1\right) + U^*(t)U(t')n(\Omega)\nonumber\\
  &+ \sum_\lambda |\kappa_\lambda|^2 \int_0^t ds \int_0^{t'} ds'\Big[U(t-s)\nonumber\\
  &\times U^*(t'-s') e^{-i\omega_\lambda(s-s')} \left(n(\omega_\lambda)+1\right)\nonumber\\
 &\,+ U^*(t-s)U(t'-s')e^{i\omega_\lambda(s-s')}n(\omega_\lambda)\Big]. \label{eq:BCF_singlePM_factor}
\end{align}
Here, $n(\omega)$ is the mean occupation number of the harmonic oscillator of frequency $\omega$, as introduced in \eref{eq:mean_occupation_number}.

For diagonal initial conditions, \eref{eq:rho_0_diag}, it is advisable to first linearly transform the operators $b$, $a_\lambda$ into the eigenbasis of the joint environment, outlined in Appendix~\ref{sec:app_bath_trafos}. We find
\begin{widetext}
\begin{align}
 \alpha(t,t')/|g|^2 &= \sum_\mu |S_{0\mu}|^2\left[U(t)U^*(t')\left(n(\tilde{\omega}_\mu)+1\right) + U^*(t)U(t')n(\tilde{\omega}_\mu) \right] 
 + \sum_{\lambda,\tau,\mu}\kappa_\lambda^*\kappa_\tau S_{\lambda\mu} S_{\tau\mu}^* \int_0^t ds \int_0^{t'} ds' \nonumber\\
 &\times \left[U(t-s)U^*(t'-s')e^{-i(\omega_\lambda s-\omega_\tau s')}\left(n(\tilde{\omega}_\mu)+1\right) + U^*(t-s)U(t'-s')e^{i(\omega_\lambda s-\omega_\tau s')} n(\tilde{\omega}_\mu)\right]\nonumber\\
 &+ i \sum_{\lambda,\mu}S_{0\mu} S_{\lambda\mu}^*\kappa_\lambda\left[U(t)\int_0^{t'} ds\,U^*(t'-s)e^{i\omega_\lambda s} \left(n(\tilde{\omega}_\mu)+1\right) +U(t')\int_0^{t} ds\,U^*(t-s)e^{i\omega_\lambda s}n(\tilde{\omega}_\mu)\right]\nonumber\\
 &- i \sum_{\lambda,\mu}S_{0\mu}^* S_{\lambda\mu}\kappa_\lambda^*\left[U^*(t')\int_0^{t} ds\,U(t-s)e^{-i\omega_\lambda s} \left(n(\tilde{\omega}_\mu)+1\right) + U^*(t)\int_0^{t'} ds\,U(t'-s)e^{-i\omega_\lambda s}n(\tilde{\omega}_\mu)\right].
 \label{eq:BCF_singlePM_diagonal}
\end{align}
\end{widetext}
Note that the above results assume a Hermitian system operator $L$ of the system-PM coupling [cf.\ \eref{eq:H_system_PM_coupling}].

The procedure allowing us to arrive at \esref{eq:BCF_singlePM_factor}{eq:BCF_singlePM_diagonal} is straightforward: We (i) derived the Heisenberg equations of motion for the PM operator, (ii) linearly transformed the time-independent ($t=0$) operators of the Heisenberg equations of motion into the operators with respect to which the initial state is diagonal (cf.\ Appendix~\ref{sec:app_bath_trafos}), and (iii) evaluated the trace of the BCF [\eref{eq:BCF_microscopical}] in this basis.

Analogously to the procedure described in Sec.~\ref{subsec:BCF_calculation}, the scheme summarized above can be generalized to, e.g., linearly-coupled chains of PMs with the last PM(s) possibly coupled to a terminating bath \cite{martinazzo2011a,woods2014}, direct coupling of PMs to independent baths \cite{ritschel2014}, or a combination of both \cite{huh2014}. Since the terminating baths are usually taken to be independent, the BCFs of the PMs can be derived by successively solving the Heisenberg equations of motion following the above approach. Starting at the PM which is coupled to the system and subsequently transforming the PM and bath operators into the basis in which the initial states are diagonal, the BCFs can be evaluated with respect to the given initial states.


\begin{thebibliography}{34}%
\makeatletter
\providecommand \@ifxundefined [1]{%
 \@ifx{#1\undefined}
}%
\providecommand \@ifnum [1]{%
 \ifnum #1\expandafter \@firstoftwo
 \else \expandafter \@secondoftwo
 \fi
}%
\providecommand \@ifx [1]{%
 \ifx #1\expandafter \@firstoftwo
 \else \expandafter \@secondoftwo
 \fi
}%
\providecommand \natexlab [1]{#1}%
\providecommand \enquote  [1]{``#1''}%
\providecommand \bibnamefont  [1]{#1}%
\providecommand \bibfnamefont [1]{#1}%
\providecommand \citenamefont [1]{#1}%
\providecommand \href@noop [0]{\@secondoftwo}%
\providecommand \href [0]{\begingroup \@sanitize@url \@href}%
\providecommand \@href[1]{\@@startlink{#1}\@@href}%
\providecommand \@@href[1]{\endgroup#1\@@endlink}%
\providecommand \@sanitize@url [0]{\catcode `\\12\catcode `\$12\catcode
  `\&12\catcode `\#12\catcode `\^12\catcode `\_12\catcode `\%12\relax}%
\providecommand \@@startlink[1]{}%
\providecommand \@@endlink[0]{}%
\providecommand \url  [0]{\begingroup\@sanitize@url \@url }%
\providecommand \@url [1]{\endgroup\@href {#1}{\urlprefix }}%
\providecommand \urlprefix  [0]{URL }%
\providecommand \Eprint [0]{\href }%
\providecommand \doibase [0]{http://dx.doi.org/}%
\providecommand \selectlanguage [0]{\@gobble}%
\providecommand \bibinfo  [0]{\@secondoftwo}%
\providecommand \bibfield  [0]{\@secondoftwo}%
\providecommand \translation [1]{[#1]}%
\providecommand \BibitemOpen [0]{}%
\providecommand \bibitemStop [0]{}%
\providecommand \bibitemNoStop [0]{.\EOS\space}%
\providecommand \EOS [0]{\spacefactor3000\relax}%
\providecommand \BibitemShut  [1]{\csname bibitem#1\endcsname}%
\let\auto@bib@innerbib\@empty
\bibitem [{\citenamefont {Weiss}(2008)}]{weiss2008}%
  \BibitemOpen
  \bibfield  {author} {\bibinfo {author} {\bibfnamefont {U.}~\bibnamefont
  {Weiss}},\ }\href@noop {} {\emph {\bibinfo {title} {Quantum Dissipative
  Systems}}}\ (\bibinfo  {publisher} {World Scientific},\ \bibinfo {year}
  {2008})\BibitemShut {NoStop}%
\bibitem [{\citenamefont {May}\ and\ \citenamefont {K{\"u}hn}(2011)}]{may2011}%
  \BibitemOpen
  \bibfield  {author} {\bibinfo {author} {\bibfnamefont {V.}~\bibnamefont
  {May}}\ and\ \bibinfo {author} {\bibfnamefont {O.}~\bibnamefont {K{\"u}hn}},\
  }\href@noop {} {\emph {\bibinfo {title} {Charge and Energy Transfer Dynamics
  in Molecular Systems}}}\ (\bibinfo  {publisher} {Wiley},\ \bibinfo {year}
  {2011})\BibitemShut {NoStop}%
\bibitem [{\citenamefont {Mukamel}(1995)}]{mukamel1995}%
  \BibitemOpen
  \bibfield  {author} {\bibinfo {author} {\bibfnamefont {S.}~\bibnamefont
  {Mukamel}},\ }\href@noop {} {\emph {\bibinfo {title} {Principles of nonlinear
  optical spectroscopy}}},\ Oxford series in optical and imaging sciences\
  (\bibinfo  {publisher} {Oxford University Press},\ \bibinfo {year}
  {1995})\BibitemShut {NoStop}%
\bibitem [{\citenamefont {Fleming}\ and\ \citenamefont
  {Cho}(1996)}]{fleming1996}%
  \BibitemOpen
  \bibfield  {author} {\bibinfo {author} {\bibfnamefont {G.~R.}\ \bibnamefont
  {Fleming}}\ and\ \bibinfo {author} {\bibfnamefont {M.}~\bibnamefont {Cho}},\
  }\href {\doibase 10.1146/annurev.physchem.47.1.109} {\bibfield  {journal}
  {\bibinfo  {journal} {Annual Review of Physical Chemistry}\ }\textbf
  {\bibinfo {volume} {47}},\ \bibinfo {pages} {109} (\bibinfo {year}
  {1996})}\BibitemShut {NoStop}%
\bibitem [{\citenamefont {Roden}\ \emph {et~al.}(2012)\citenamefont {Roden},
  \citenamefont {Strunz}, \citenamefont {Whaley},\ and\ \citenamefont
  {Eisfeld}}]{roden2012}%
  \BibitemOpen
  \bibfield  {author} {\bibinfo {author} {\bibfnamefont {J.}~\bibnamefont
  {Roden}}, \bibinfo {author} {\bibfnamefont {W.~T.}\ \bibnamefont {Strunz}},
  \bibinfo {author} {\bibfnamefont {K.~B.}\ \bibnamefont {Whaley}}, \ and\
  \bibinfo {author} {\bibfnamefont {A.}~\bibnamefont {Eisfeld}},\ }\href
  {\doibase http://dx.doi.org/10.1063/1.4765329} {\bibfield  {journal}
  {\bibinfo  {journal} {The Journal of Chemical Physics}\ }\textbf {\bibinfo
  {volume} {137}},\ \bibinfo {eid} {204110} (\bibinfo {year}
  {2012})}\BibitemShut {NoStop}%
\bibitem [{\citenamefont {Di\'osi}\ \emph {et~al.}(1998)\citenamefont
  {Di\'osi}, \citenamefont {Gisin},\ and\ \citenamefont {Strunz}}]{diosi1998}%
  \BibitemOpen
  \bibfield  {author} {\bibinfo {author} {\bibfnamefont {L.}~\bibnamefont
  {Di\'osi}}, \bibinfo {author} {\bibfnamefont {N.}~\bibnamefont {Gisin}}, \
  and\ \bibinfo {author} {\bibfnamefont {W.~T.}\ \bibnamefont {Strunz}},\
  }\href {\doibase 10.1103/PhysRevA.58.1699} {\bibfield  {journal} {\bibinfo
  {journal} {Phys. Rev. A}\ }\textbf {\bibinfo {volume} {58}},\ \bibinfo
  {pages} {1699} (\bibinfo {year} {1998})}\BibitemShut {NoStop}%
\bibitem [{\citenamefont {Roden}\ \emph {et~al.}(2011)\citenamefont {Roden},
  \citenamefont {Strunz},\ and\ \citenamefont {Eisfeld}}]{roden2011}%
  \BibitemOpen
  \bibfield  {author} {\bibinfo {author} {\bibfnamefont {J.}~\bibnamefont
  {Roden}}, \bibinfo {author} {\bibfnamefont {W.~T.}\ \bibnamefont {Strunz}}, \
  and\ \bibinfo {author} {\bibfnamefont {A.}~\bibnamefont {Eisfeld}},\ }\href
  {\doibase http://dx.doi.org/10.1063/1.3512979} {\bibfield  {journal}
  {\bibinfo  {journal} {The Journal of Chemical Physics}\ }\textbf {\bibinfo
  {volume} {134}},\ \bibinfo {eid} {034902} (\bibinfo {year}
  {2011})}\BibitemShut {NoStop}%
\bibitem [{\citenamefont {Suess}\ \emph {et~al.}(2014)\citenamefont {Suess},
  \citenamefont {Eisfeld},\ and\ \citenamefont {Strunz}}]{suess2014}%
  \BibitemOpen
  \bibfield  {author} {\bibinfo {author} {\bibfnamefont {D.}~\bibnamefont
  {Suess}}, \bibinfo {author} {\bibfnamefont {A.}~\bibnamefont {Eisfeld}}, \
  and\ \bibinfo {author} {\bibfnamefont {W.~T.}\ \bibnamefont {Strunz}},\
  }\href {\doibase 10.1103/PhysRevLett.113.150403} {\bibfield  {journal}
  {\bibinfo  {journal} {Phys. Rev. Lett.}\ }\textbf {\bibinfo {volume} {113}},\
  \bibinfo {pages} {150403} (\bibinfo {year} {2014})}\BibitemShut {NoStop}%
\bibitem [{\citenamefont {Luoma}\ \emph {et~al.}(2014)\citenamefont {Luoma},
  \citenamefont {Haikka},\ and\ \citenamefont {Piilo}}]{luoma2014}%
  \BibitemOpen
  \bibfield  {author} {\bibinfo {author} {\bibfnamefont {K.}~\bibnamefont
  {Luoma}}, \bibinfo {author} {\bibfnamefont {P.}~\bibnamefont {Haikka}}, \
  and\ \bibinfo {author} {\bibfnamefont {J.}~\bibnamefont {Piilo}},\ }\href
  {\doibase 10.1103/PhysRevA.90.054101} {\bibfield  {journal} {\bibinfo
  {journal} {Phys. Rev. A}\ }\textbf {\bibinfo {volume} {90}},\ \bibinfo
  {pages} {054101} (\bibinfo {year} {2014})}\BibitemShut {NoStop}%
\bibitem [{\citenamefont {Shabani}\ \emph {et~al.}(2014)\citenamefont
  {Shabani}, \citenamefont {Roden},\ and\ \citenamefont
  {Whaley}}]{shabani2014}%
  \BibitemOpen
  \bibfield  {author} {\bibinfo {author} {\bibfnamefont {A.}~\bibnamefont
  {Shabani}}, \bibinfo {author} {\bibfnamefont {J.}~\bibnamefont {Roden}}, \
  and\ \bibinfo {author} {\bibfnamefont {K.~B.}\ \bibnamefont {Whaley}},\
  }\href {\doibase 10.1103/PhysRevLett.112.113601} {\bibfield  {journal}
  {\bibinfo  {journal} {Phys. Rev. Lett.}\ }\textbf {\bibinfo {volume} {112}},\
  \bibinfo {pages} {113601} (\bibinfo {year} {2014})}\BibitemShut {NoStop}%
\bibitem [{\citenamefont {Garg}\ \emph {et~al.}(1985)\citenamefont {Garg},
  \citenamefont {Onuchic},\ and\ \citenamefont {Ambegaokar}}]{garg1985}%
  \BibitemOpen
  \bibfield  {author} {\bibinfo {author} {\bibfnamefont {A.}~\bibnamefont
  {Garg}}, \bibinfo {author} {\bibfnamefont {J.~N.}\ \bibnamefont {Onuchic}}, \
  and\ \bibinfo {author} {\bibfnamefont {V.}~\bibnamefont {Ambegaokar}},\
  }\href {\doibase http://dx.doi.org/10.1063/1.449017} {\bibfield  {journal}
  {\bibinfo  {journal} {The Journal of Chemical Physics}\ }\textbf {\bibinfo
  {volume} {83}},\ \bibinfo {pages} {4491} (\bibinfo {year}
  {1985})}\BibitemShut {NoStop}%
\bibitem [{\citenamefont {Hughes}\ \emph {et~al.}(2009)\citenamefont {Hughes},
  \citenamefont {Christ},\ and\ \citenamefont {Burghardt}}]{hughes2009}%
  \BibitemOpen
  \bibfield  {author} {\bibinfo {author} {\bibfnamefont {K.~H.}\ \bibnamefont
  {Hughes}}, \bibinfo {author} {\bibfnamefont {C.~D.}\ \bibnamefont {Christ}},
  \ and\ \bibinfo {author} {\bibfnamefont {I.}~\bibnamefont {Burghardt}},\
  }\href {\doibase http://dx.doi.org/10.1063/1.3159671} {\bibfield  {journal}
  {\bibinfo  {journal} {The Journal of Chemical Physics}\ }\textbf {\bibinfo
  {volume} {131}},\ \bibinfo {eid} {024109} (\bibinfo {year}
  {2009})}\BibitemShut {NoStop}%
\bibitem [{\citenamefont {Martinazzo}\ \emph
  {et~al.}(2011{\natexlab{a}})\citenamefont {Martinazzo}, \citenamefont
  {Vacchini}, \citenamefont {Hughes},\ and\ \citenamefont
  {Burghardt}}]{martinazzo2011}%
  \BibitemOpen
  \bibfield  {author} {\bibinfo {author} {\bibfnamefont {R.}~\bibnamefont
  {Martinazzo}}, \bibinfo {author} {\bibfnamefont {B.}~\bibnamefont
  {Vacchini}}, \bibinfo {author} {\bibfnamefont {K.~H.}\ \bibnamefont
  {Hughes}}, \ and\ \bibinfo {author} {\bibfnamefont {I.}~\bibnamefont
  {Burghardt}},\ }\href {\doibase http://dx.doi.org/10.1063/1.3532408}
  {\bibfield  {journal} {\bibinfo  {journal} {The Journal of Chemical Physics}\
  }\textbf {\bibinfo {volume} {134}},\ \bibinfo {eid} {011101} (\bibinfo {year}
  {2011}{\natexlab{a}})}\BibitemShut {NoStop}%
\bibitem [{\citenamefont {Breuer}\ and\ \citenamefont
  {Petruccione}(2002)}]{breuer2002}%
  \BibitemOpen
  \bibfield  {author} {\bibinfo {author} {\bibfnamefont {H.}~\bibnamefont
  {Breuer}}\ and\ \bibinfo {author} {\bibfnamefont {F.}~\bibnamefont
  {Petruccione}},\ }\href@noop {} {\emph {\bibinfo {title} {The Theory of Open
  Quantum Systems}}}\ (\bibinfo  {publisher} {Oxford University Press},\
  \bibinfo {address} {Oxford},\ \bibinfo {year} {2002})\BibitemShut {NoStop}%
\bibitem [{\citenamefont {Strunz}(1996)}]{strunz1996}%
  \BibitemOpen
  \bibfield  {author} {\bibinfo {author} {\bibfnamefont {W.~T.}\ \bibnamefont
  {Strunz}},\ }\href {\doibase http://dx.doi.org/10.1016/S0375-9601(96)00805-5}
  {\bibfield  {journal} {\bibinfo  {journal} {Physics Letters A}\ }\textbf
  {\bibinfo {volume} {224}},\ \bibinfo {pages} {25 } (\bibinfo {year}
  {1996})}\BibitemShut {NoStop}%
\bibitem [{\citenamefont {Ritschel}\ \emph {et~al.}(2015)\citenamefont
  {Ritschel}, \citenamefont {Suess}, \citenamefont {M\"obius}, \citenamefont
  {Strunz},\ and\ \citenamefont {Eisfeld}}]{ritschel2015}%
  \BibitemOpen
  \bibfield  {author} {\bibinfo {author} {\bibfnamefont {G.}~\bibnamefont
  {Ritschel}}, \bibinfo {author} {\bibfnamefont {D.}~\bibnamefont {Suess}},
  \bibinfo {author} {\bibfnamefont {S.}~\bibnamefont {M\"obius}}, \bibinfo
  {author} {\bibfnamefont {W.~T.}\ \bibnamefont {Strunz}}, \ and\ \bibinfo
  {author} {\bibfnamefont {A.}~\bibnamefont {Eisfeld}},\ }\href {\doibase
  http://dx.doi.org/10.1063/1.4905327} {\bibfield  {journal} {\bibinfo
  {journal} {The Journal of Chemical Physics}\ }\textbf {\bibinfo {volume}
  {142}},\ \bibinfo {eid} {034115} (\bibinfo {year} {2015})}\BibitemShut
  {NoStop}%
\bibitem [{\citenamefont {Martinazzo}\ \emph
  {et~al.}(2011{\natexlab{b}})\citenamefont {Martinazzo}, \citenamefont
  {Hughes},\ and\ \citenamefont {Burghardt}}]{martinazzo2011a}%
  \BibitemOpen
  \bibfield  {author} {\bibinfo {author} {\bibfnamefont {R.}~\bibnamefont
  {Martinazzo}}, \bibinfo {author} {\bibfnamefont {K.}~\bibnamefont {Hughes}},
  \ and\ \bibinfo {author} {\bibfnamefont {I.}~\bibnamefont {Burghardt}},\
  }\href {\doibase 10.1103/PhysRevE.84.030102} {\bibfield  {journal} {\bibinfo
  {journal} {Phys. Rev. E}\ }\textbf {\bibinfo {volume} {84}},\ \bibinfo
  {pages} {030102} (\bibinfo {year} {2011}{\natexlab{b}})}\BibitemShut
  {NoStop}%
\bibitem [{\citenamefont {Woods}\ \emph {et~al.}(2014)\citenamefont {Woods},
  \citenamefont {Groux}, \citenamefont {Chin}, \citenamefont {Huelga},\ and\
  \citenamefont {Plenio}}]{woods2014}%
  \BibitemOpen
  \bibfield  {author} {\bibinfo {author} {\bibfnamefont {M.~P.}\ \bibnamefont
  {Woods}}, \bibinfo {author} {\bibfnamefont {R.}~\bibnamefont {Groux}},
  \bibinfo {author} {\bibfnamefont {A.~W.}\ \bibnamefont {Chin}}, \bibinfo
  {author} {\bibfnamefont {S.~F.}\ \bibnamefont {Huelga}}, \ and\ \bibinfo
  {author} {\bibfnamefont {M.~B.}\ \bibnamefont {Plenio}},\ }\href {\doibase
  http://dx.doi.org/10.1063/1.4866769} {\bibfield  {journal} {\bibinfo
  {journal} {Journal of Mathematical Physics}\ }\textbf {\bibinfo {volume}
  {55}},\ \bibinfo {eid} {032101} (\bibinfo {year} {2014})}\BibitemShut
  {NoStop}%
\bibitem [{\citenamefont {Huh}\ \emph {et~al.}(2014)\citenamefont {Huh},
  \citenamefont {Mostame}, \citenamefont {Fujita}, \citenamefont {Yung},\ and\
  \citenamefont {Aspuru-Guzik}}]{huh2014}%
  \BibitemOpen
  \bibfield  {author} {\bibinfo {author} {\bibfnamefont {J.}~\bibnamefont
  {Huh}}, \bibinfo {author} {\bibfnamefont {S.}~\bibnamefont {Mostame}},
  \bibinfo {author} {\bibfnamefont {T.}~\bibnamefont {Fujita}}, \bibinfo
  {author} {\bibfnamefont {M.-H.}\ \bibnamefont {Yung}}, \ and\ \bibinfo
  {author} {\bibfnamefont {A.}~\bibnamefont {Aspuru-Guzik}},\ }\href
  {http://stacks.iop.org/1367-2630/16/i=12/a=123008} {\bibfield  {journal}
  {\bibinfo  {journal} {New Journal of Physics}\ }\textbf {\bibinfo {volume}
  {16}},\ \bibinfo {pages} {123008} (\bibinfo {year} {2014})}\BibitemShut
  {NoStop}%
\bibitem [{\citenamefont {Garraway}(1997)}]{garraway1997}%
  \BibitemOpen
  \bibfield  {author} {\bibinfo {author} {\bibfnamefont {B.~M.}\ \bibnamefont
  {Garraway}},\ }\href {\doibase 10.1103/PhysRevA.55.2290} {\bibfield
  {journal} {\bibinfo  {journal} {Phys. Rev. A}\ }\textbf {\bibinfo {volume}
  {55}},\ \bibinfo {pages} {2290} (\bibinfo {year} {1997})}\BibitemShut
  {NoStop}%
\bibitem [{\citenamefont {Kubo}\ \emph {et~al.}(1991)\citenamefont {Kubo},
  \citenamefont {Toda}, \citenamefont {Sait{\=o}},\ and\ \citenamefont
  {Hashitsume}}]{kubo1991}%
  \BibitemOpen
  \bibfield  {author} {\bibinfo {author} {\bibfnamefont {R.}~\bibnamefont
  {Kubo}}, \bibinfo {author} {\bibfnamefont {M.}~\bibnamefont {Toda}}, \bibinfo
  {author} {\bibfnamefont {N.}~\bibnamefont {Sait{\=o}}}, \ and\ \bibinfo
  {author} {\bibfnamefont {N.}~\bibnamefont {Hashitsume}},\ }\href@noop {}
  {\emph {\bibinfo {title} {Statistical Physics II: Nonequilibrium Statistical
  Mechanics}}}\ (\bibinfo  {publisher} {Springer Berlin Heidelberg},\ \bibinfo
  {year} {1991})\BibitemShut {NoStop}%
\bibitem [{\citenamefont {Chaudhry}\ and\ \citenamefont
  {Gong}(2013)}]{chaudhry2013}%
  \BibitemOpen
  \bibfield  {author} {\bibinfo {author} {\bibfnamefont {A.~Z.}\ \bibnamefont
  {Chaudhry}}\ and\ \bibinfo {author} {\bibfnamefont {J.}~\bibnamefont
  {Gong}},\ }\href {\doibase 10.1103/PhysRevA.88.052107} {\bibfield  {journal}
  {\bibinfo  {journal} {Phys. Rev. A}\ }\textbf {\bibinfo {volume} {88}},\
  \bibinfo {pages} {052107} (\bibinfo {year} {2013})}\BibitemShut {NoStop}%
\bibitem [{\citenamefont {Yang}\ \emph {et~al.}(2014)\citenamefont {Yang},
  \citenamefont {An}, \citenamefont {Luo}, \citenamefont {Li},\ and\
  \citenamefont {Oh}}]{yang2014}%
  \BibitemOpen
  \bibfield  {author} {\bibinfo {author} {\bibfnamefont {C.-J.}\ \bibnamefont
  {Yang}}, \bibinfo {author} {\bibfnamefont {J.-H.}\ \bibnamefont {An}},
  \bibinfo {author} {\bibfnamefont {H.-G.}\ \bibnamefont {Luo}}, \bibinfo
  {author} {\bibfnamefont {Y.}~\bibnamefont {Li}}, \ and\ \bibinfo {author}
  {\bibfnamefont {C.~H.}\ \bibnamefont {Oh}},\ }\href {\doibase
  10.1103/PhysRevE.90.022122} {\bibfield  {journal} {\bibinfo  {journal} {Phys.
  Rev. E}\ }\textbf {\bibinfo {volume} {90}},\ \bibinfo {pages} {022122}
  (\bibinfo {year} {2014})}\BibitemShut {NoStop}%
\bibitem [{\citenamefont {Iles-Smith}\ \emph {et~al.}(2014)\citenamefont
  {Iles-Smith}, \citenamefont {Lambert},\ and\ \citenamefont
  {Nazir}}]{iles-smith2014}%
  \BibitemOpen
  \bibfield  {author} {\bibinfo {author} {\bibfnamefont {J.}~\bibnamefont
  {Iles-Smith}}, \bibinfo {author} {\bibfnamefont {N.}~\bibnamefont {Lambert}},
  \ and\ \bibinfo {author} {\bibfnamefont {A.}~\bibnamefont {Nazir}},\ }\href
  {\doibase 10.1103/PhysRevA.90.032114} {\bibfield  {journal} {\bibinfo
  {journal} {Phys. Rev. A}\ }\textbf {\bibinfo {volume} {90}},\ \bibinfo
  {pages} {032114} (\bibinfo {year} {2014})}\BibitemShut {NoStop}%
\bibitem [{\citenamefont {van Kampen}(2004)}]{kampen2004}%
  \BibitemOpen
  \bibfield  {author} {\bibinfo {author} {\bibfnamefont {N.}~\bibnamefont {van
  Kampen}},\ }\href {\doibase 10.1023/B:JOSS.0000022383.06086.6c} {\bibfield
  {journal} {\bibinfo  {journal} {Journal of Statistical Physics}\ }\textbf
  {\bibinfo {volume} {115}},\ \bibinfo {pages} {1057} (\bibinfo {year}
  {2004})}\BibitemShut {NoStop}%
\bibitem [{\citenamefont {Ambegaokar}(2007)}]{ambegaokar2007}%
  \BibitemOpen
  \bibfield  {author} {\bibinfo {author} {\bibfnamefont {V.}~\bibnamefont
  {Ambegaokar}},\ }\href {\doibase 10.1002/andp.200610236} {\bibfield
  {journal} {\bibinfo  {journal} {Annalen der Physik}\ }\textbf {\bibinfo
  {volume} {16}},\ \bibinfo {pages} {319} (\bibinfo {year} {2007})}\BibitemShut
  {NoStop}%
\bibitem [{\citenamefont {Ritschel}\ and\ \citenamefont
  {Eisfeld}(2014)}]{ritschel2014}%
  \BibitemOpen
  \bibfield  {author} {\bibinfo {author} {\bibfnamefont {G.}~\bibnamefont
  {Ritschel}}\ and\ \bibinfo {author} {\bibfnamefont {A.}~\bibnamefont
  {Eisfeld}},\ }\href {\doibase http://dx.doi.org/10.1063/1.4893931} {\bibfield
   {journal} {\bibinfo  {journal} {The Journal of Chemical Physics}\ }\textbf
  {\bibinfo {volume} {141}},\ \bibinfo {eid} {094101} (\bibinfo {year}
  {2014})}\BibitemShut {NoStop}%
\bibitem [{\citenamefont {Hakim}\ and\ \citenamefont
  {Ambegaokar}(1985)}]{hakim1985}%
  \BibitemOpen
  \bibfield  {author} {\bibinfo {author} {\bibfnamefont {V.}~\bibnamefont
  {Hakim}}\ and\ \bibinfo {author} {\bibfnamefont {V.}~\bibnamefont
  {Ambegaokar}},\ }\href {\doibase 10.1103/PhysRevA.32.423} {\bibfield
  {journal} {\bibinfo  {journal} {Phys. Rev. A}\ }\textbf {\bibinfo {volume}
  {32}},\ \bibinfo {pages} {423} (\bibinfo {year} {1985})}\BibitemShut
  {NoStop}%
\bibitem [{\citenamefont {Pollak}\ \emph {et~al.}(2008)\citenamefont {Pollak},
  \citenamefont {Shao},\ and\ \citenamefont {Zhang}}]{pollak2008}%
  \BibitemOpen
  \bibfield  {author} {\bibinfo {author} {\bibfnamefont {E.}~\bibnamefont
  {Pollak}}, \bibinfo {author} {\bibfnamefont {J.}~\bibnamefont {Shao}}, \ and\
  \bibinfo {author} {\bibfnamefont {D.~H.}\ \bibnamefont {Zhang}},\ }\href
  {\doibase 10.1103/PhysRevE.77.021107} {\bibfield  {journal} {\bibinfo
  {journal} {Phys. Rev. E}\ }\textbf {\bibinfo {volume} {77}},\ \bibinfo
  {pages} {021107} (\bibinfo {year} {2008})}\BibitemShut {NoStop}%
\bibitem [{\citenamefont {Suárez}\ \emph {et~al.}(1992)\citenamefont
  {Suárez}, \citenamefont {Silbey},\ and\ \citenamefont
  {Oppenheim}}]{suarez1992}%
  \BibitemOpen
  \bibfield  {author} {\bibinfo {author} {\bibfnamefont {A.}~\bibnamefont
  {Suárez}}, \bibinfo {author} {\bibfnamefont {R.}~\bibnamefont {Silbey}}, \
  and\ \bibinfo {author} {\bibfnamefont {I.}~\bibnamefont {Oppenheim}},\ }\href
  {\doibase http://dx.doi.org/10.1063/1.463831} {\bibfield  {journal} {\bibinfo
   {journal} {The Journal of Chemical Physics}\ }\textbf {\bibinfo {volume}
  {97}},\ \bibinfo {pages} {5101} (\bibinfo {year} {1992})}\BibitemShut
  {NoStop}%
\bibitem [{\citenamefont {Morozov}\ \emph {et~al.}(2012)\citenamefont
  {Morozov}, \citenamefont {Mathey},\ and\ \citenamefont
  {R\"opke}}]{morozov2012}%
  \BibitemOpen
  \bibfield  {author} {\bibinfo {author} {\bibfnamefont {V.~G.}\ \bibnamefont
  {Morozov}}, \bibinfo {author} {\bibfnamefont {S.}~\bibnamefont {Mathey}}, \
  and\ \bibinfo {author} {\bibfnamefont {G.}~\bibnamefont {R\"opke}},\ }\href
  {\doibase 10.1103/PhysRevA.85.022101} {\bibfield  {journal} {\bibinfo
  {journal} {Phys. Rev. A}\ }\textbf {\bibinfo {volume} {85}},\ \bibinfo
  {pages} {022101} (\bibinfo {year} {2012})}\BibitemShut {NoStop}%
\bibitem [{\citenamefont {Valleau}\ \emph {et~al.}(2012)\citenamefont
  {Valleau}, \citenamefont {Eisfeld},\ and\ \citenamefont
  {Aspuru-Guzik}}]{valleau2012}%
  \BibitemOpen
  \bibfield  {author} {\bibinfo {author} {\bibfnamefont {S.}~\bibnamefont
  {Valleau}}, \bibinfo {author} {\bibfnamefont {A.}~\bibnamefont {Eisfeld}}, \
  and\ \bibinfo {author} {\bibfnamefont {A.}~\bibnamefont {Aspuru-Guzik}},\
  }\href {\doibase http://dx.doi.org/10.1063/1.4769079} {\bibfield  {journal}
  {\bibinfo  {journal} {The Journal of Chemical Physics}\ }\textbf {\bibinfo
  {volume} {137}},\ \bibinfo {eid} {224103} (\bibinfo {year}
  {2012})}\BibitemShut {NoStop}%
\bibitem [{\citenamefont {Tan}\ and\ \citenamefont {Zhang}(2011)}]{tan2011}%
  \BibitemOpen
  \bibfield  {author} {\bibinfo {author} {\bibfnamefont {H.-T.}\ \bibnamefont
  {Tan}}\ and\ \bibinfo {author} {\bibfnamefont {W.-M.}\ \bibnamefont
  {Zhang}},\ }\href {\doibase 10.1103/PhysRevA.83.032102} {\bibfield  {journal}
  {\bibinfo  {journal} {Phys. Rev. A}\ }\textbf {\bibinfo {volume} {83}},\
  \bibinfo {pages} {032102} (\bibinfo {year} {2011})}\BibitemShut {NoStop}%
\bibitem [{\citenamefont {Zhang}\ \emph {et~al.}(2012)\citenamefont {Zhang},
  \citenamefont {Lo}, \citenamefont {Xiong}, \citenamefont {Tu},\ and\
  \citenamefont {Nori}}]{zhang2012}%
  \BibitemOpen
  \bibfield  {author} {\bibinfo {author} {\bibfnamefont {W.-M.}\ \bibnamefont
  {Zhang}}, \bibinfo {author} {\bibfnamefont {P.-Y.}\ \bibnamefont {Lo}},
  \bibinfo {author} {\bibfnamefont {H.-N.}\ \bibnamefont {Xiong}}, \bibinfo
  {author} {\bibfnamefont {M.~W.-Y.}\ \bibnamefont {Tu}}, \ and\ \bibinfo
  {author} {\bibfnamefont {F.}~\bibnamefont {Nori}},\ }\href {\doibase
  10.1103/PhysRevLett.109.170402} {\bibfield  {journal} {\bibinfo  {journal}
  {Phys. Rev. Lett.}\ }\textbf {\bibinfo {volume} {109}},\ \bibinfo {pages}
  {170402} (\bibinfo {year} {2012})}\BibitemShut {NoStop}%
\end{thebibliography}
%

\end{document}